\documentclass{jfm}
\usepackage{hyperref}
\hypersetup{
colorlinks = true,
urlcolor   = blue,
citecolor  = blue,
linkcolor  = blue,
}
\usepackage{amsmath}
\usepackage{amssymb}
\usepackage{graphicx}
\usepackage{epstopdf, epsfig}
\usepackage{psfrag}
\usepackage{subcaption}
\usepackage{cleveref}
\usepackage{xcolor}
\usepackage{mathtools}
\usepackage{mathrsfs}
\usepackage{esint}
\usepackage{tikz}
\usetikzlibrary{decorations.pathmorphing}

\newtheorem{lemma}{Lemma}

\newcommand{\bs}[1]{\boldsymbol{#1}}
\newcommand*{\QEDB}{\hfill\ensuremath{\square}}

\def\Xint#1{\mathchoice
{\XXint\displaystyle\textstyle{#1}}%
{\XXint\textstyle\scriptstyle{#1}}%
{\XXint\scriptstyle\scriptscriptstyle{#1}}%
{\XXint\scriptscriptstyle\scriptscriptstyle{#1}}%
\!\int}
\def\XXint#1#2#3{{\setbox0=\hbox{$#1{#2#3}{\int}$ }
\vcenter{\hbox{$#2#3$ }}\kern-.6\wd0}}

\def\dashint{\Xint-}

\newtheorem{innercustomgeneric}{\customgenericname}
\providecommand{\customgenericname}{}
\newcommand{\newcustomtheorem}[2]{%
\newenvironment{#1}[1]
{%
\renewcommand\customgenericname{#2}%
\renewcommand\theinnercustomgeneric{##1}%
\innercustomgeneric
}
{\endinnercustomgeneric}
}

\newcustomtheorem{customthm}{Theorem}
\newcustomtheorem{customlemma}{Lemma}

\shorttitle{Bound on heat transport in internally heated convection}

\title{Analytical bounds on the heat transport in internally heated convection}

\author{Anuj Kumar\aff{1}
\corresp{\email{akumar43@ucsc.edu}}, Ali Arslan\aff{2}
\corresp{\email{a.arslan18@imperial.ac.uk}}, 
Giovanni Fantuzzi\aff{2},
John Craske\aff{3} and Andrew Wynn\aff{2}
}

\affiliation{\aff{1}Department of Applied Mathematics,   University of California, Santa Cruz, CA 95064, USA \aff{2}Department of Aeronautics, Imperial College London, SW7 2AZ, UK
\aff{3}Department of Civil and Environmental Engineering, Imperial College London, SW7 2AZ, UK }

\begin{document}

\maketitle

\begin{abstract}
%% Gio's slightly shortened version
We obtain an analytical bound on the mean vertical convective heat flux $\langle w T \rangle$ between two parallel boundaries driven by uniform internal heating. We consider two configurations, one with both boundaries held at the same constant temperature, and the other one with a top boundary held at constant temperature and a perfectly insulating bottom boundary. For the first configuration, Arslan \textit{et al.} (\href{https://doi.org/10.1017/jfm.2021.360}{\textit{J. Fluid Mech.} 919:A15, 2021}) recently provided numerical evidence that Rayleigh-number-dependent corrections to the only known rigorous bound $\langle w T \rangle \leq 1/2$ may be provable if the classical background method is augmented with a minimum principle stating that the fluid's temperature is no smaller than that of the top boundary. Here, we confirm this fact rigorously for both configurations by proving bounds on $\langle wT \rangle$ that approach $1/2$ exponentially from below as the Rayleigh number is increased. The key to obtaining these bounds are inner boundary layers in the background fields with a particular inverse-power scaling, which can be controlled in the spectral constraint using Hardy and Rellich inequalities. These allow for qualitative improvements in the analysis not available to standard constructions.
\end{abstract}


%just to see the structure while working on the paper
% \tableofcontents

\section{Introduction}
\label{sec:intro}
Convection driven by buoyancy is abundant in geophysical and astrophysical flows, from atmospheric convection driving ocean currents to solar convection transporting heat in stars. The prototypical setup for studying these flows is that of Rayleigh--B\'enard convection, where flow in a layer of fluid is driven by the temperature differential across the boundaries. In reality, convection in many natural or engineering situations is at least partially driven by an internal heating source. Examples include convection in the Earth's mantle due to radiogenic heat \citep{davies1992mantle, schubert2001mantle,mulyukova2020}, convection in radiative planet atmospheres \citep{seager2010exoplanet, pierrehumbert2010principles, guervilly2019turbulent}, and engineering flows where exothermic chemical or nuclear reactions drive the convection \citep{tran2009effective}. Gaining insights into these physical and practical scenarios requires a thorough understanding of internally heated (IH) convection, and yet studies in this direction are relatively few.

Following the early investigations by \cite{roberts1967convection} and \cite{tritton1975internally}, research into IH convection has recently gained renewed momentum through computational analysis \citep{goluskin2012convection, goluskin2015internally, goluskin2015penetrative} and experiments \citep{lepot2018radiative, bouillaut2019transition,limare2019convection,limare2021transient}. However, a comprehensive understanding of flows driven by internal heating is far from complete and the behaviour of such flows in the limiting regime of extreme heating remains unknown. 

\begin{figure}
    \centering
    \begin{tabular}{lc}
    \begin{subfigure}{0.5\textwidth}
    \centering
    \includegraphics[scale = 0.45]{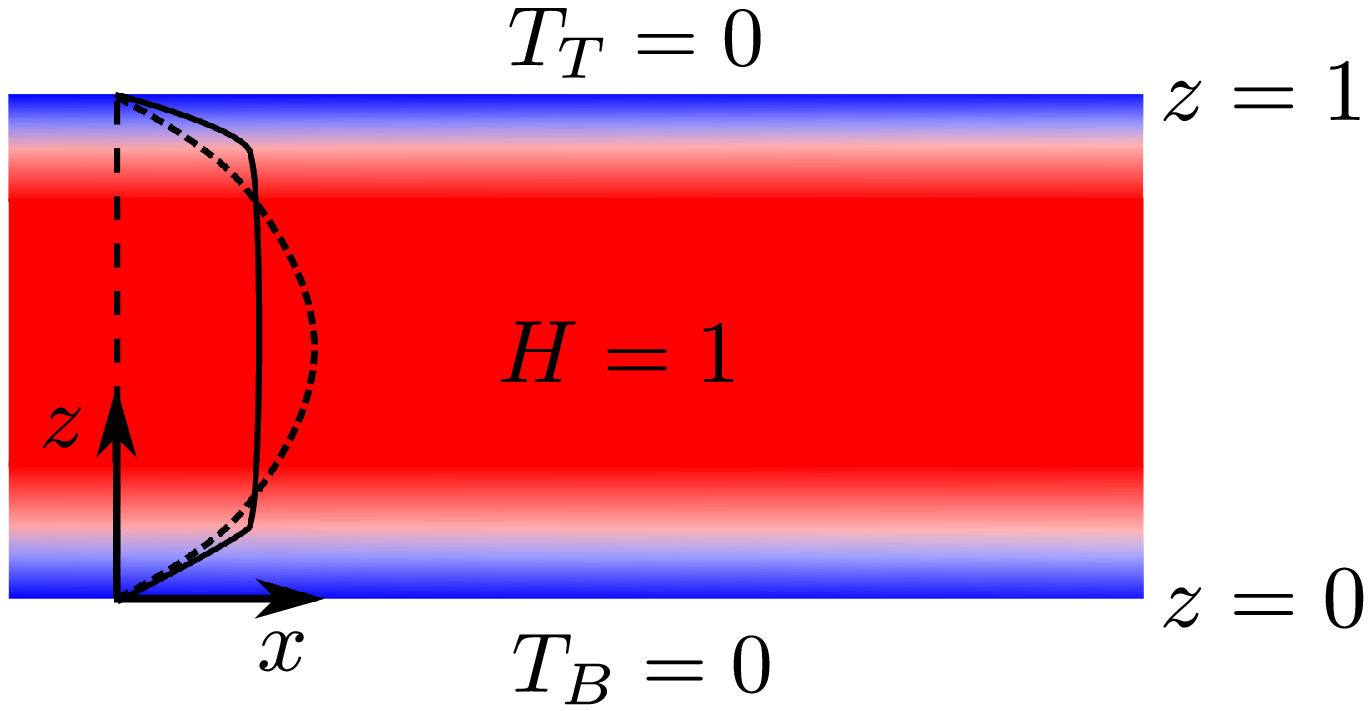}
    \caption{}
    \end{subfigure} &
    \begin{subfigure}{0.5\textwidth}
    \centering
    \includegraphics[scale = 0.45]{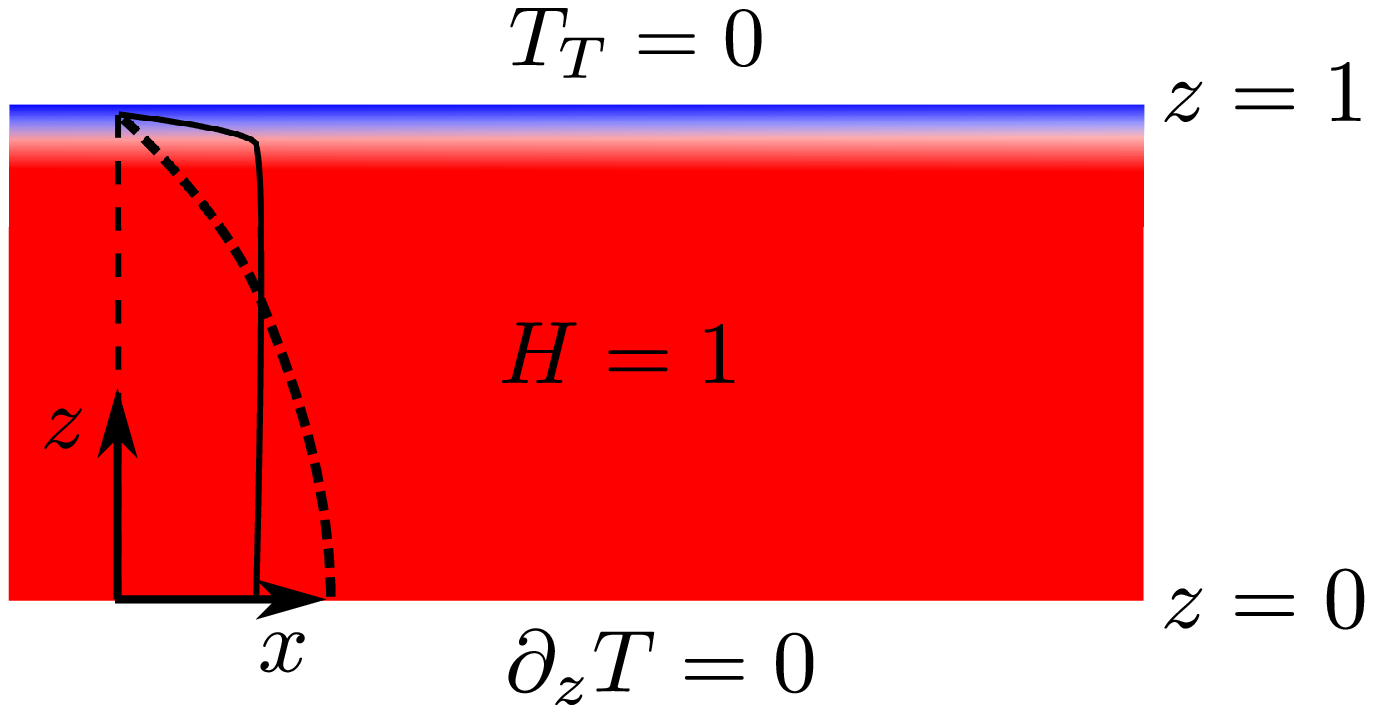}
    \caption{}
    \end{subfigure}
    \end{tabular}
    \caption{The two configurations considered in this paper. (a) IH1: Isothermal boundaries, (b) IH3: Isothermal top boundary and insulating bottom boundary. In both configurations the heating is uniform, so the non-dimensional thermal source term is $H = 1$.  Dashed lines show the temperature profiles in the pure conduction state, while solid lines sketch the temporally- and horizontally-averaged temperature profiles in a typical turbulent state (also shown using the color plot).}
     \label{Flow configuration: IH1 and IH3 figures}
\end{figure}

Here, we probe this regime using rigorous upper bounding theory. Specifically, we bound the mean vertical convective heat flux in two configurations of IH convection, one where the fluid is bounded between horizontal plates held at the same temperature and one where the bottom plate is replaced by a perfect insulator. These two configurations, which we refer to as IH1 and IH3 following the terminology introduced by \cite{goluskin2016internally}, are illustrated schematically in panels (a) and (b) of figure~\ref{Flow configuration: IH1 and IH3 figures}. %In both cases, a minimum principle ensures that the temperature of the fluid cannot be larger than that of the isothermal top boundary.

The mean vertical convective heat flux $\langle wT \rangle$, where $w$ and $T$ are the nondimensional vertical velocity and temperature and angled brackets denote space-time averages, has a slightly different physical interpretation in the two configurations. For the IH1 case, $\langle w T \rangle$  is related to the asymmetry in the heat fluxes $\mathcal{F}_T$ and $\mathcal{F}_B$ through the top and the bottom boundaries. Specifically, space-time averaging the dimensionless transport equation for temperature (see~\eqref{e:heat-ns} in \S\ref{s:setup}) multiplied by the wall-normal coordinate $z$ yields
\begin{equation}
\mathcal{F}_T = \frac{1}{2} + \langle w T \rangle, \qquad \mathcal{F}_B = \frac{1}{2} - \langle w T \rangle.
\end{equation}
In the purely conductive state, the heat generated inside the domain leaves equally between the two boundaries, hence $\mathcal{F}_T = \mathcal{F}_B = 1/2$. In the convective state, instead, the asymmetry of buoyancy combines with the uniform heat source to create boundary layers with different characteristics near the top and bottom boundaries, as illustrated in figure~\ref{Flow configuration: IH1 and IH3 figures}(a). The bottom boundary layer is stably stratified, whereas the top boundary layer is unstably stratified. Convective heat transport ($\langle wT\rangle > 0$) makes the top boundary layer thinner than the bottom one, so in any convective state one has $\mathcal{F}_T > \mathcal{F}_B$. Since the boundary temperature is fixed and the fluid is internally heated, one also expects the boundary flux $\mathcal{F}_B$ to remain non-negative, meaning that heat can escape from the bottom boundary but not enter through it. This fact can be proved rigorously (\citealp[Appendix~A.1]{goluskin2012convection}; \citealp[Appendix~A]{arslan2021bounds}) and translates into the following upper bounds on the vertical heat transport \citep{goluskin2012convection}:
\begin{equation}
\langle w T \rangle \leq \frac{1}{2} \quad \text{in IH1}.
\label{Introduction: hypothesis IH1}
\end{equation}

For the IH3 configuration, instead, the mean vertical flux $\langle w T \rangle$ is related to the difference of the horizontally-averaged temperature between the top $\overline{T}_T$ and the bottom wall $\overline{T}_B$. Indeed, upon multiplying the dimensionless evolution equation for the temperature (see~\eqref{e:heat-ns} in \S\ref{s:setup}) with the wall-normal coordinate $z$ and space-time averaging one obtains
\begin{eqnarray}
\langle w T \rangle  = \overline{T}_T -\overline{T}_B + \frac12.
\end{eqnarray}
The isothermal boundary condition implies that the temperature $T_T$ at the top boundary is in fact constant, so $\overline{T}_T = T_T$, and we take it be zero without loss of generality in our nondimensionalization. Since the nondimensional internal heating rate is positive, one expects the mean bottom temperature $\overline{T}_B$ to be non-negative. As before, this fact can be proved rigorously and results in the upper bound \cite[Chapter~1]{goluskin2016internally}
\begin{equation}
\langle w T \rangle \leq \frac{1}{2}\quad \text{in IH3}.
\label{Introduction: hypothesis IH3}
\end{equation}

For the IH1 configuration, \cite{arslan2021bounds} recently proved that $\langle w T \rangle \leq 2^{-21/5}R^{1/5}$, where $R$ is a nondimensional parameter that measures the strength of the internal heating and may be interpreted as a Rayleigh number. This result, which is independent of the Prandtl number $\Pran$, fails to improve the uniform bound in~\eqref{Introduction: hypothesis IH1} for $R > 2^{16}=65536$. However, numerical evidence by the same authors suggests that an upper bound on  $\langle w T \rangle$ approaching $1/2$ from below monotonically as $R$ is increased may be provable when the background method by Doering \& Constantin \citep{PhysRevLett.69.1648,PhysRevE.49.4087,PhysRevE.53.5957,PhysRevE.51.3192} is augmented with a minimum principle stating that the fluid's temperature cannot be smaller than that the top boundary. Unfortunately, they also provided a rather tantalizing proof that such a bound cannot be obtained using typical analytical constructions. 

In this paper we overcome this barrier and show that $R$-dependent bounds on $\langle wT \rangle$ strictly smaller than $1/2$ can be obtained analytically not only in the IH1 case, but also for the IH3 configuration. Precisely, we prove that
\begin{subequations} \label{intro bounds}
\begin{align} 
\label{e:ih1-bound-intro}
&&&&\langle w T \rangle &\leq \frac{1}{2} -  c_1 R^{\frac{1}{5}} \exp \left(-c_2 R^{\frac{3}{5}}\right) &\text{in IH1,} &&&&\\
&&&&\langle w T \rangle &\leq \frac{1}{2} -  \frac{c_3}{R^{\frac{1}{5}}} \exp \left(-c_4 R^{\frac{3}{5}}\right) &\text{in IH3,}&&&&
\label{e:ih3-bound-intro}
\end{align} 
\end{subequations}
where $c_1, c_2, c_3$ and $c_4$ are constants (independent of both $R$ and $\Pran$). To establish these results, we formulate a bounding principle for $\langle wT \rangle$ using the auxiliary functional method \citep{chernyshenko2014polynomial, fantuzzi2016bounds, tobasco2018optimal, chernyshenko2017relationship}. This method is a generalization of the background method of Doering and Constantin, which has successfully been applied to several fluid dynamical problems \citep{PhysRevLett.69.1648, PhysRevE.51.3192, PhysRevE.53.5957, caulfield2001maximal, tang2004bounds, whitehead2011ultimate, goluskin2016bounds, fantuzzi2018boundsA, fantuzzi2018boundsB, kumar2020bound, kumar2020pressure, fan2021three, arslan2021bounds2, arslan2021bounds, kumar2021optimal}. The auxiliary functional method, as implemented in this paper, also has an equivalent formulation using the background method.

The novelty aspects in our arguments are the use of a background temperature field with a lower boundary layer growing as $z^{-1}$, motivated by the numerical results by \citet{arslan2021bounds}, and the application of Hardy inequalities (IH1) and Rellich inequalities (IH3). Such inequalities have already been employed to prove bounds on convective flows at infinite Prandtl number \citep{doering2006bounds,whitehead2011internal} but, to the best of our knowledge, their use at finite Prandtl number is new.

The rest of this work is organized as follows. We start by describing the problem setup in \S 2.  In \S 3, we apply the auxiliary function method formulate upper bounding principles for $\langle wT \rangle$ in both IH1 and IH3 configurations. We then prove the upper bound~\eqref{e:ih1-bound-intro} in \S 4 and the upper bound~\eqref{e:ih3-bound-intro} in \S 5. Finally, \S 6, discusses our method of proof, compares our results with available phenomenological theories, and offers concluding remarks.

\section{Problem setup}\label{s:setup}
We consider the flow of a Newtonian fluid of density $\rho$, viscosity $\nu$ and thermal diffusivity $\kappa$ driven by buoyancy forces resulting from internal heating. The fluid is confined between two horizontal no-slip plates with a gap of width $d$ and the heat is produced at a constant volumetric rate of $H^\ast / c_p$, where $c_p$ is the fluid's heat capacity. %We call $H^\ast(\boldsymbol{x}, t)$ the thermal source term and, in this paper, we restrict ourselves to the case where it is constant in time and space, i.e., $H^\ast(\boldsymbol{x}, t) \equiv H^\ast$ . 
We consider the two configurations sketched in figure \ref{Flow configuration: IH1 and IH3 figures}, one where both plates are kept a constant temperature $T_0^\ast$ (IH1) and one where the top plate is kept at a constant temperature $T_0^\ast$ while the bottom plate is insulating (IH3). 
% Following the terminology from \citet{goluskin2016internally}, we call the former configuration IH1 and the later one IH3. 

We assume that the fluid properties are a weak function of the temperature and use the Naiver--Stokes equations under the Boussinesq approximation to model the problem. 
Various justifications have been put forward for the Boussinesq approximation; see, for example, \citet{spiegel1960boussinesq} and \citet{rajagopal1996oberbeck}. In their non-dimensional form, the governing equations are
\begin{subequations}
\begin{align}
\label{eq:continuity}
\bnabla \bcdot \boldsymbol{u} &= 0, \\
\label{e:momentum-ns}
\partial_t \boldsymbol{u} + \boldsymbol{u} \bcdot \bnabla \boldsymbol{u}+ \bnabla p &=  \Pran \nabla^2 \boldsymbol{u} + \Pran R T \boldsymbol{e}_z, \\
\label{e:heat-ns}
\partial_t T + \boldsymbol{u} \bcdot \bnabla T &= \bnabla^2 T + 1,
\end{align}
\label{Flow configuration: governing equations}
\end{subequations}
where we have used the following non-dimensionalization for the variables:
\begin{eqnarray}
\bs{x} = \frac{\bs{x}^\ast}{d}, \quad t = \frac{t^\ast}{d^2/ \kappa}, \quad \bs{u} = \frac{\bs{u}^\ast}{\kappa/d},  \quad p = \frac{p^\ast - p_0}{\rho \kappa^2/d^2}, \quad T = \frac{T^\ast - T_0^\ast}{d^2 H / \kappa}.
\label{Flow configuration: nondim}
\end{eqnarray}
Here, $\bs{x}$, $t$, $\bs{u}$, $p$ and $T$ denote the non-dimensional position, time, velocity, pressure and temperature, respectively, whereas  $p_0$ is the dimensional hydrostatic ambient pressure. The quantities with a star in superscript are dimensional. The non-dimensional governing parameters of the flow are the Prandtl number and the Rayleigh number, given by
\begin{equation}
\Pran = \frac{\nu}{\kappa} \qquad\text{and}\qquad R = \frac{g \alpha d^5 H}{\nu \kappa},
\end{equation}
where $\alpha$ is the coefficient of thermal expansion. 

We use the Cartesian coordinates $\bs{x} = (x, y, z)$ and place the origin of the coordinate system at the bottom plate. The $z$-direction points vertically upward and the $x$ and $y$ directions are horizontal. In this coordinate system, we write the velocity vector as $\bs{u} = (u, v, w)$ where $u$, $v$ and $w$ are the velocity components in the $x$, $y$ and $z$ directions respectively. In this coordinate system, the boundary conditions at the top and bottom plates for velocity and temperature can be written as
\begin{subequations}
\label{Flow configuration: boundary conditions}
\begin{eqnarray}
\bs{u}(x, y, 0, t) = \bs{u}(x, y, 1, t) = \bs{0}, && \\
T(x, y, 0, t) = T(x, y, 1, t) = 0 && \quad \text{for IH1,} \\
\partial_z T(x, y, 0, t) = T(x, y, 1, t) = 0 && \quad \text{for IH3.} 
\end{eqnarray}
\end{subequations}
We further assume that the fluid layer is periodic in the horizontal directions $x$ and $y$ with length $L_x$ and $L_y$, meaning that the domain of interest is $\Omega = \mathbb{T}_{[0, L_x]} \times \mathbb{T}_{[0, L_y]} \times [0, 1]$.

Throughout the paper, spatial averages, long-time horizontal averages and long-time volume averages will be denoted, respectively, by
%We use the following notation for the spatial and long-time-horizontal and -volume averages,
\begin{subequations}
\begin{eqnarray}
&& \dashint_{\Omega} [\;\cdot\;] \; {\rm d}\bs{x} = \frac{1}{L_x L_y} \int_{0}^{1} \int_{0}^{L_y} \int_{0}^{L_x} [\;\cdot\;] \; \textrm{d}x \textrm{d}y \textrm{d}z, \\
&& \overline{[\;\cdot\;]} = \lim_{\tau \to \infty} \frac{1}{\tau L_x L_y} \int_{0}^\tau \int_{0}^{L_y} \int_{0}^{L_x} [\;\cdot\;] \; \textrm{d}x \textrm{d}y \textrm{d}t, \\
&& \langle[\;\cdot\;]\rangle = \lim_{\tau \to \infty} \frac{1}{\tau} \int_{0}^\tau \dashint_{\Omega} [\;\cdot\;] \; \textrm{d}\bs{x} \textrm{d}t. 
\end{eqnarray}
\label{Flow configuration: average notation}
\end{subequations}
% The goal of this paper is to obtain bounds on the non-dimensional heat transfer $\langle w T \rangle$ in both IH1 and IH3 configurations as a function of the Rayleigh number $R$ in the presence of a minimum principle.

%% Anuj's version of sec 3
\section{The auxiliary functional method}
\label{The auxiliary functions method formulation}
A bound on the mean vertical heat flux can be derived using the auxiliary function method. The formulation of the method given here is very similar to the one given by \citet{arslan2021bounds} for isothermal boundaries, but we repeat it to make the paper self-contained and highlight the changes required when the lower boundary is insulating.

Let $\mathcal{V}\{\bs{u}, T\}$ be a functional that is uniformly bounded in time along solutions $\bs{u}(t)$ and $T(t)$ of the governing equations (\ref{Flow configuration: governing equations}a-c).Further, let $\mathcal{L}\{\bs{u}, T\}$ be the Lie derivative of $\mathcal{V}\{\bs{u}, T\}$, meaning a functional such that
\begin{eqnarray}
\mathcal{L}\{\bs{u}(t), T(t)\} = \frac{{\rm d}}{{\rm d} t} \mathcal{V}\{\bs{u}(t), T(t)\}
\label{Formulation: functional L}
\end{eqnarray}
when $\bs{u}(t)$ and $T(t)$ solve the governing equations. Then, a simple calculation shows that the long-time average of $\mathcal{L}\{\bs{u}(t), T(t)\}$ vanishes and we can rewrite the mean vertical heat flux as
\begin{eqnarray}
\langle w T \rangle && = \lim_{\tau \to \infty} \frac{1}{\tau} \int_{0}^{\tau} \left[ \dashint_{\Omega} w T  \,{\rm d}\bs{x} + \mathcal{L}\{\bs{u}(t), T(t)\} \right] \; {\rm d}t, \nonumber \\
&& = B + \lim_{\tau \to \infty} \frac{1}{\tau} \int_{0}^{\tau} \left[ \dashint_{\Omega} w T  \,{\rm d}\bs{x} + \mathcal{L}\{\bs{u}(t), T(t)\} - B \right] \; {\rm d}t.
\label{Formulation: intermediate bound 1}
\end{eqnarray}
If the functional $\mathcal{V}$ can be chosen such that %the term in the square brackets on the right hand side of (\ref{Formulation: intermediate bound 1}) is always non-positive, that is
\begin{eqnarray}
\mathcal{S}^\ast\{\bs{u}, T\} \coloneqq \dashint_{\Omega} w T  \,{\rm d}\bs{x} + \mathcal{L} \{\bs{u}, T\} - B  \leq 0
\label{Formulation: intermediate bound 2 requirement}
\end{eqnarray}
for any solution of the governing equations, then it follows that $\langle w T \rangle \leq B$. Of course, it is intractable to impose \eqref{Formulation: intermediate bound 2 requirement} only over the set of solutions of the governing equation, because they are not known explicitly. However, to obtain a (possibly conservative) bound it suffices to enforce the stronger condition that \eqref{Formulation: intermediate bound 2 requirement} holds for all pairs of divergence-free velocity fields $\bs{u}$ and temperature fields $T$ that satisfy the boundary conditions (\ref{Flow configuration: boundary conditions}a-c). %Of course, such fields contain all solutions to the governing equations. 
%  \begin{eqnarray}
%  \langle w T \rangle \leq B.
%  \label{Formulation: intermediate bound 2}
%  \end{eqnarray}

%In theory, we should be able to choose a $\mathcal{V}$ (subject to technicalities) such that the bound is sharp 
% i.e., there exists a solution of the governing equations (\ref{Flow configuration: governing equations}a-c) for which $\langle w T \rangle = B$ 
%\citep{tobasco2018optimal, rosa2020optimal}. However, the expression of such a functional is currently unknown. The most general quadratic function that is known to work is given as
Following \citet{arslan2021bounds}, we choose the functional $\mathcal{V}$ to be
\begin{eqnarray}
\mathcal{V} \{\bs{u}, T\} = \dashint_{\Omega} \left[\frac{a}{2 Pr R} |\bs{u}|^2 + b |T|^2 - (\psi(z) + z - 1) T \right] \; {\rm d} \bs{x}.
\label{Formulation: general expression of V}
\end{eqnarray} 
%where the nonnegative scalars $a$ and $b$, the function $\phi$ and the vector field $\boldsymbol{\chi}$ are to be optimised to obtain the best possible bound. In relation to the background method from \cite{chernyshenko2017relationship}, the profile $\frac{\phi}{b}$ is the background temperature field, $\frac{Pr R}{a}\boldsymbol{\chi}$ the background velocity and $a$ and $b$ are balance parameters. We need not impose any boundary conditions on $\phi$ and $\boldsymbol{\chi}$ at this stage and, as discussed by \cite{fantuzzi2021background}, the symmetries of the governing equations in \eqref{Flow configuration: governing equations} can be combined with the incompressibility and no-slip boundary conditions to conclude that there is no loss of generality in setting $\boldsymbol{\chi} =\bs{0}$ and $\phi(\boldsymbol{x})= \phi(z)$.  Following \citet{arslan2021bounds}, we choose 
%\begin{eqnarray}
%\phi(z) = \psi(z) + z - 1
%\end{eqnarray}
Differentiating this functional in time along solutions of the governing equations, followed by standard integrations by parts using the divergence-free and boundary conditions, yields an expression for $\mathcal{L}\{\bs{u},T\}$ that can be substituted into \eqref{Formulation: intermediate bound 2 requirement} to obtain
\enlargethispage{\baselineskip}
% Using the choice (\ref{Formulation: general expression of V}) together with the definition (\ref{Formulation: functional L}) and the time derivatives of various quantities obtained from the governing equations, the condition (\ref{Formulation: intermediate bound 2 requirement}) becomes
\begin{multline}
\mathcal{S}^\ast\{\bs{u}, T\} = \dashint_{\Omega} \left[ \frac{a}{R}|\bnabla \bs{u}|^2 + b |\bnabla T|^2 - (a - \psi^\prime) w T + (bz - \psi^\prime) \frac{\partial T}{\partial z} + \psi \right] {\rm d} \bs{x} \\
+ T(0) - T(1) + \psi(1) \left. \overline{\frac{\partial T}{\partial z}}\right|_{z = 1} - (\psi(0) - 1) \left. \overline{\frac{\partial T}{\partial z}} \right|_{z = 0} + B - \frac{1}{2} \geq 0.
\label{Formulation: functional S tilde}
\end{multline}
This inequality needs to be satisfied for all $\bs{u}$ and $T$ satisfying \eqref{eq:continuity}, (\ref{Flow configuration: boundary conditions}a) and either (\ref{Flow configuration: boundary conditions}b) for IH1 or (\ref{Flow configuration: boundary conditions}c) for IH3. 
%  \begin{subequations}
%  \begin{eqnarray}
%  \mathcal{H}_{1} \coloneqq \{(\bs{u}, T) \; | \; \bs{u} \text{ satisfies } (\ref{Flow configuration: governing equations}a) \text{ and } (\ref{Flow configuration: boundary conditions}a), T \text{ satisfies } (\ref{Flow configuration: boundary conditions}b)\} \quad \text{for IH1} \\
%  \mathcal{H}_{3} \coloneqq \{(\bs{u}, T) \; | \; \bs{u} \text{ satisfies } (\ref{Flow configuration: governing equations}a) \text{ and } (\ref{Flow configuration: boundary conditions}a), T \text{ satisfies } (\ref{Flow configuration: boundary conditions}c)\} \quad \text{for IH3.}
%  \end{eqnarray}
%  \end{subequations}
% for (\ref{Formulation: intermediate bound 2}) to hold. 

A crucial improvement to the best upper bound $B$ implied by~\eqref{Formulation: functional S tilde} can be achieved by imposing the minimum principle, which says that $T\geq 0 $ at all times if it is so initially, and that any negative component decays exponentially quickly \citep{arslan2021bounds}. We may therefore restrict the attention to nonnegative temperature fields, thereby relaxing inequality~\eqref{Formulation: functional S tilde}.
%   However, it is difficult use this information ($T \geq 0$) explicitly into our proof. Instead, 
As explained by \cite{arslan2021bounds}, the constraint can be enforced with the help of a nondecreasing Lagrange multiplier function $q(z)$ by adding the term
\begin{eqnarray}
\dashint_{\Omega} q^\prime(z) T {\rm d} \bs{x}
\label{Formulation: Lagrange multiplier}
\end{eqnarray}
to the right-hand side of~\eqref{Formulation: functional S tilde}. Integrating by parts and rearranging leads to the weaker constraint
\begin{eqnarray}
\mathcal{S}\{\bs{u}, T\} \coloneqq \mathcal{S}^\ast\{\bs{u}, T\} + \dashint_{\Omega} q(z) \frac{\partial T}{\partial z} {\rm d}\bs{x} + q(0) T(0) - q(1) T(1) \geq 0,
\label{Formulation: functional S not so early}
\end{eqnarray}
and the best upper bound on $\langle wT \rangle$ implied by this inequality is
\begin{align}
\langle w T \rangle \leq \inf_{B,\psi(z),q(z),a,b} \bigg\{B: \quad &q(z)\, \text{non-decreasing}, \nonumber\\[-2ex]
&\mathcal{S}\{\bs{u}, T\} \geq 0 \quad \forall(\boldsymbol{u},T)~\text{satisfying~\eqref{eq:continuity} and \eqref{Flow configuration: boundary conditions}}
\bigg\}.
\label{e:bound_with_positivity_statement}
\end{align} 
Moreover, since no derivatives of  the Lagrange multiplier $q(z)$ appear in inequality~\eqref{Formulation: functional S not so early}, one can perform the optimization over nondecreasing Lagrange multipliers that are not necessarily differentiable everywhere and may even be discontinuous. A rigorous justification of this statement is given by \cite{arslan2021bounds}.

%We can observe that setting $q(z)=-1$ causes the integral in (\ref{Formulation: functional S not so early}) to vanish and thus become identical to the previous spectral constraint (\ref{Formulation: functional S tilde}).

To prove an explicit rigorous bound on $\langle wT \rangle$, it is convenient to replace inequality~\eqref{Formulation: functional S not so early} with a stronger condition that is more amenable to analytical treatment. To achieve this, we introduce the following Fourier series decomposition of the variables in the $x$ and $y$ directions: 
\begin{eqnarray}
\begin{bmatrix}
\bs{u}(\bs{x}) \\
T(\bs{x})
\end{bmatrix} =
\sum_{\bs{k} \in K} 
\begin{bmatrix}
\hat{\bs{u}}_{\bs{k}}(z) \\
\hat{T}_{\bs{k}}(z) 
\end{bmatrix} e^{i k_x x + i k_y y},
\label{Formulation: Fourier decomposition}
\end{eqnarray}
where
\begin{eqnarray}
K \equiv \left\{ (k_x, k_y) = \left. \left(\frac{2 m \upi}{L_x}, \frac{2 n \upi}{L_y}\right) \; \right|  (m, n) \in \mathbb{Z}^2 \right\}.
\end{eqnarray}
%Using this Fourier decomposition, we derive a stricter but simplified spectral condition, which is more amenable to analytical treatment.
Since $\bs{u}$ and $T$ in (\ref{Formulation: Fourier decomposition})  are real-valued, the Fourier expansion coefficients satisfy $\hat{w}_{\bs{k}}^\ast = \hat{w}_{-\bs{k}}$ and $\hat{T}_{\bs{k}}^\ast = \hat{T}_{-\bs{k}}$ for all $\bs{k} \in K$, subject to the boundary conditions
\begin{subequations}\label{e:Fourier-bc}
\begin{gather}
    % \hat{u}_{\boldsymbol{k}}(0) = \hat{u}_{\boldsymbol{k}}(1)=\hat{v}_{\boldsymbol{k}}(0) = \hat{v}_{\boldsymbol{k}}(1)=0,\\
    \label{e:Fourier-bc-wk}
    \hat{w}_{\boldsymbol{k}}(0) = \hat{w}_{\boldsymbol{k}}'(0) = \hat{w}_{\boldsymbol{k}}(1) = \hat{w}_{\boldsymbol{k}}'(1)= 0,\\
    \hat{T}_{\boldsymbol{k}}(0)=\hat{T}_{\boldsymbol{k}}(1) = 0, \qquad \textrm{IH1,}          \label{e:Fourier-bc-Tk_2}\\
     \hat{T}_{\boldsymbol{k}}'(0)=\hat{T}_{\boldsymbol{k}}(1) = 0, \qquad \textrm{IH3.}
    \label{e:Fourier-bc-Tk_3}
\end{gather}
\end{subequations}

Substituting (\ref{Formulation: Fourier decomposition}) in  (\ref{Formulation: functional S not so early}), using the incompressiblity condition on $\bs{u}$, applying the inequality of arithmetic and geometric means (AM--GM inequality), and dropping positive terms in $\hat{u}_{\boldsymbol{k}}$ and $\hat{v}_{\boldsymbol{k}}$, we can estimate
\begin{equation}\label{e:S-lower-bound-fourier}
\mathcal{S}\{\bs{u}, T\} \geq \mathcal{S}_{\bs{0}} \{\hat{T}_{\bs{0}}\} + \sum_{\bs{k} \neq \bs{0}} \mathcal{S}_{\bs{k}} \{\hat{w}_{\bs{k}}, \hat{T}_{\bs{k}}\} ,
\end{equation}
where
\begin{multline}
\mathcal{S}_{\bs{0}} \{\hat{T}_{\bs{0}}\} \coloneqq \int_0^1 \left[b |\hat{T}_{\bs{0}}^{\prime}|^2 + (b z - \psi^\prime + q) \hat{T}_{\bs{0}}^\prime + \psi\right] {\rm d}z + (q(0) + 1) \hat{T}_{\bs{0}}(0)   \\
- (q(1) + 1) \hat{T}_{\bs{0}}(1) + \psi(1) \hat{T}_{\bs{0}}^\prime(1) - (\psi(0) - 1) \hat{T}^\prime_{\bs{0}}(0) + B - \frac{1}{2},
\label{Formulation: expression S0}
\end{multline}
and 
\begin{multline}
\mathcal{S}_{\bs{k}} \{\hat{w}_{\bs{k}}, \hat{T}_{\bs{k}}\} \coloneqq \int_{0}^{1} \left[\frac{a}{R}\left(\frac{1}{k^2}|\hat{w}_{\bs{k}}^{\prime \prime}|^2 + 2 |\hat{w}_{\bs{k}}^{\prime}|^2 + k^2 |\hat{w}_{\bs{k}}|^2 \right) \right. \qquad \qquad \\ \left. + b |\hat{T}_{\bs{k}}^{\prime}|^2 + b k^2 |\hat{T}_{\bs{k}}|^2 - (a - \psi^\prime) \hat{w}_{\bs{k}}\hat{T}_{\bs{k}}^{\ast}\right] {\rm d}z.
\label{Formulation: expression Sk}
\end{multline}
In the last expression, $k = \sqrt{k_x^2 + k_y^2}$. 

To establish inequality~\eqref{Formulation: functional S not so early}, therefore, it suffices to check the nonnegativity of the right-hand side of~\eqref{e:S-lower-bound-fourier}. As all the different Fourier modes $\hat{w}_{\bs{k}}$ and $\hat{T}_{\bs{k}}$ can be chosen independently, this requires
$\mathcal{S}_{\bs{k}} \{\hat{w}_{\bs{k}}, \hat{T}_{\bs{k}}\} + \mathcal{S}_{-\bs{k}} \{\hat{w}_{-\bs{k}}, \hat{T}_{-\bs{k}}\} \geq 0$ for all wavevectors $\bs{k} \in K$, which in turn holds true if and only if 
$\mathcal{S}_{\bs{k}} \{\operatorname{Re}\{\hat{w}_{\bs{k}}\}, \operatorname{Re}\{\hat{T}_{\bs{k}}\}\} \geq 0$ and $\mathcal{S}_{\bs{k}} \{\operatorname{Im}\{\hat{w}_{\bs{k}}\}, \operatorname{Im}\{\hat{T}_{\bs{k}}\}\} \geq 0$ for all wavevectors $\bs{k} \in K$.
% \begin{eqnarray}
% \mathcal{S}_{\bs{k}} \{\hat{w}_{\bs{k}}, \hat{T}_{\bs{k}}\} + \mathcal{S}_{-\bs{k}} \{\hat{w}_{-\bs{k}}, \hat{T}_{-\bs{k}}\} \geq 0 \quad \forall \bs{k} \in K,
% \end{eqnarray}
% which holds true if one can show
% \begin{eqnarray}
% \mathcal{S}_{\bs{k}} \{\operatorname{Re}\{\hat{w}_{\bs{k}}\}, \operatorname{Re}\{\hat{T}_{\bs{k}}\}\} \geq 0 \quad \text{and} \quad \mathcal{S}_{\bs{k}} \{\operatorname{Im}\{\hat{w}_{\bs{k}}\}, \operatorname{Im}\{\hat{T}_{\bs{k}}\}\} \geq 0 \quad \forall \bs{k} \in K.
% \end{eqnarray}
This, combined with the fact that the real and imaginary parts of  $\hat{w}_{\bs{k}}$ and $\hat{T}_{\bs{k}}$  can be chosen independently, implies that we may take $\hat{w}_{\bs{k}}$ and $\hat{T}_{\bs{k}}$ to be real-valued without loss of generality and impose
\begin{subequations}
\begin{align}
\mathcal{S}_{\bs{0}} \{\hat{T}_{\bs{0}}\} &\geq 0, \\
\mathcal{S}_{\bs{k}}\{\hat{w}_{\bs{k}},\hat{T}_{\bs{k}}\} &\geq 0 \quad \forall \bs{k} \in K, \, \bs{k}\neq \bs{0}.
\label{Formulation: spectral constraint intermediate}
\end{align}
\end{subequations}
%We refer to the second set of conditions as the \textit{spectral constraints}.

From the nonnegativity condition on $\mathcal{S}_{\bs{0}} \{\hat{T}_{\bs{0}}\}$, it is possible to extract the bound $B$ explicitly. First of all, the nonnegativity of $\mathcal{S}_{\bs{0}} \{\hat{T}_{\bs{0}}\}$ requires
\begin{subequations}
\label{Formulation: spectral BC psi and q}
\begin{align}
&&&&&& \psi(0) &= 1, &\psi(1) &= 0 &&\text{for IH1}, &&&&&&\\
&&&&&& q(0) &= -1, & \psi(1) &= 0  &&\text{for IH3},&&&&&&
\end{align}
\end{subequations}
otherwise it is possible to choose a profile $\hat{T}_{\bs{0}}(z)$ that is non-zero only near the boundaries and for which $\mathcal{S}_{\bs{0}} \{\hat{T}_{\bs{0}}\} \leq 0$. With these simplifications, one can write
\begin{eqnarray}
\mathcal{S}_{\bs{0}} \{\hat{T}_{\bs{0}}\} = \int_0^1 \left[ \sqrt{b} \hat{T}_{\bs{0}}^{\prime} + \frac{(b z - \psi^\prime + q)}{2 \sqrt{b}}\right]^2 {\rm d}z + B - \frac{1}{4 b} \int_0^1 (b z - \psi^\prime + q)^2 {\rm d}z\nonumber \\  + \int_0^1 \psi(z) {\rm d}z - \frac{1}{2}.
\end{eqnarray}
Therefore, $\mathcal{S}_{\bs{0}} \{\hat{T}_{\bs{0}}\}$ is nonnegative if we choose $B$ to cancel the negative and sign-indefinite terms.
%  \begin{eqnarray}
%  B = \inf_{a, b, \psi(z), q(z)} \left\{ \frac{1}{2} + \frac{1}{4 b} \int_0^1 (b z - \psi^\prime + q)^2 dz - \int_0^1 \psi(z) dz \right\}.
%  \label{Formulation: good value of B}
%  \end{eqnarray}
After gathering \eqref{e:bound_with_positivity_statement}, \eqref{Formulation: Fourier decomposition}, \eqref{e:Fourier-bc}, \eqref{Formulation: spectral constraint intermediate} and \eqref{Formulation: spectral BC psi and q} we conclude that
\begin{eqnarray}
\langle w T \rangle \leq  \inf_{a, b, \psi(z), q(z)} \left\{ \frac{1}{2} + \frac{1}{4 b} \int_0^1 (b z - \psi^\prime + q)^2 {\rm d}z - \int_0^1 \psi(z) {\rm d}z \right\},
\label{Formulation: bound on wT}
\end{eqnarray}
provided
\begin{subequations}
\begin{eqnarray}
&& q(z) \text{ is a nondecreasing function}, \\
&& \psi(0) = 1, \quad \psi(1) = 0 \quad \text{for IH1}, \\
&& q(0) = -1, \quad \psi(1) = 0  \quad \text{for IH3}, \\
&& \mathcal{S}_{\bs{k}} \{\hat{w}_{\bs{k}}, \hat{T}_{\bs{k}}\} \geq 0 \quad \forall\hat{w}_{\bs{k}}, \hat{T}_{\bs{k}}:\eqref{e:Fourier-bc}, \;  \forall\bs{k}\neq \bs{0}
\end{eqnarray}
\label{Formulation: bound on wT constraints}
\end{subequations}

Explicit constructions for which the right-hand side of \eqref{Formulation: bound on wT} is strictly less than 1/2 at all Rayleigh numbers are given in \S\ref{Bound on heat flux in IH1 configuration lala} and \S\ref{Bound on heat flux in IH3 configuration lala} for the IH1 and IH3 configurations, respectively. First, however, we summarize our proof strategy to explain the intuition behind our constructions. From (\ref{Formulation: bound on wT}), we see that the competition between the second term (which is always positive) and the third term will decide if $\langle w T \rangle$ can be less than $1/2$ as long as we are able to enforce that $\mathcal{S}_{\bs{k}} \{\hat{w}_{\bs{k}}, \hat{T}_{\bs{k}}\} \geq 0$. For previous studies using the background method, the standard approach has been to choose a profile $\psi(z)$ that is linear in boundary layers near the walls, whereas in the bulk region $\psi(z)$ is chosen such that the sign indefinite term in $\mathcal{S}_{\bs{k}}$ is zero. Unfortunately, in the present case, for a profile of $\psi(z)$ which is linear in the boundary layers, we are unable to show that the magnitude of the second term in (\ref{Formulation: bound on wT}) is smaller than the third term unless we violate the constraint~\eqref{Formulation: spectral constraint intermediate}. However, if we use a $z^{-1}$ profile in $\psi(z)$ in the outer layer of a two-layer lower boundary layer---a choice inspired by numerical computation from \citet{arslan2021bounds}---we gain an extra factor of a logarithm in the integral of $\psi$. This makes it possible to show that sum of second and third term in (\ref{Formulation: bound on wT}) is negative without violating $\mathcal{S}_{\bs{k}} \{\hat{w}_{\bs{k}}, \hat{T}_{\bs{k}}\} \geq 0$. This observation relies on the application of the following Hardy and Rellich inequalities, proofs of which are provided for completeness in Appendix \ref{Proof of Hardy and Rellich inequalities}. 
\begin{lemma}[Hardy inequality]\label{th:hardy}
Let $f: [0, \infty) \to \mathbb{R}$ be a function such that $f, f' \in L^2(0, \infty)$ and such that $f(0) = 0$. Then, for any $\epsilon > 0$ and any $\alpha \geq 0$,
\begin{eqnarray}
\int_{0}^{\alpha} \frac{|f|^2}{(z + \epsilon)^2} {\rm d}z \leq 4 \int_{0}^{\alpha} |f'|^2 {\rm d} z.
\end{eqnarray}
\end{lemma}
\begin{lemma}[Rellich inequality]\label{th:rellich}
Let $f: [0, \infty) \to \mathbb{R}$ be function such that $f, f', f'' \in L^2(0, \infty)$ and such that $f(0) = f'(0) = 0$. Then, for any $\epsilon > 0$ and any $\alpha \geq 0$,
\begin{eqnarray}
\int_{0}^{\alpha} \frac{|f|^2}{(z + \epsilon)^4} {\rm d}z \leq \frac{16}{9} \int_{0}^{\alpha} |f''|^2 {\rm d} z.
\end{eqnarray}
\end{lemma}
We now present detailed proofs of the main results. Our emphasis is on the steps necessary to obtain an $R$-dependent bound on $\langle w T \rangle$, and we do not attempt to optimize the constants appearing in our estimates.% the coefficients preceding the $R$ correction are not considered.

\section{Bound on heat flux in IH1 configuration}
\label{Bound on heat flux in IH1 configuration lala}

%In this section, we obtain an explicit bound on the heat flux $\langle w T \rangle$ as a function of the Rayleigh number  using the auxiliary function method formulation from the last section. 

\begin{figure}
\centering
\begin{tabular}{lc}
\begin{subfigure}{0.5\textwidth}
\centering
\includegraphics[scale = 0.4]{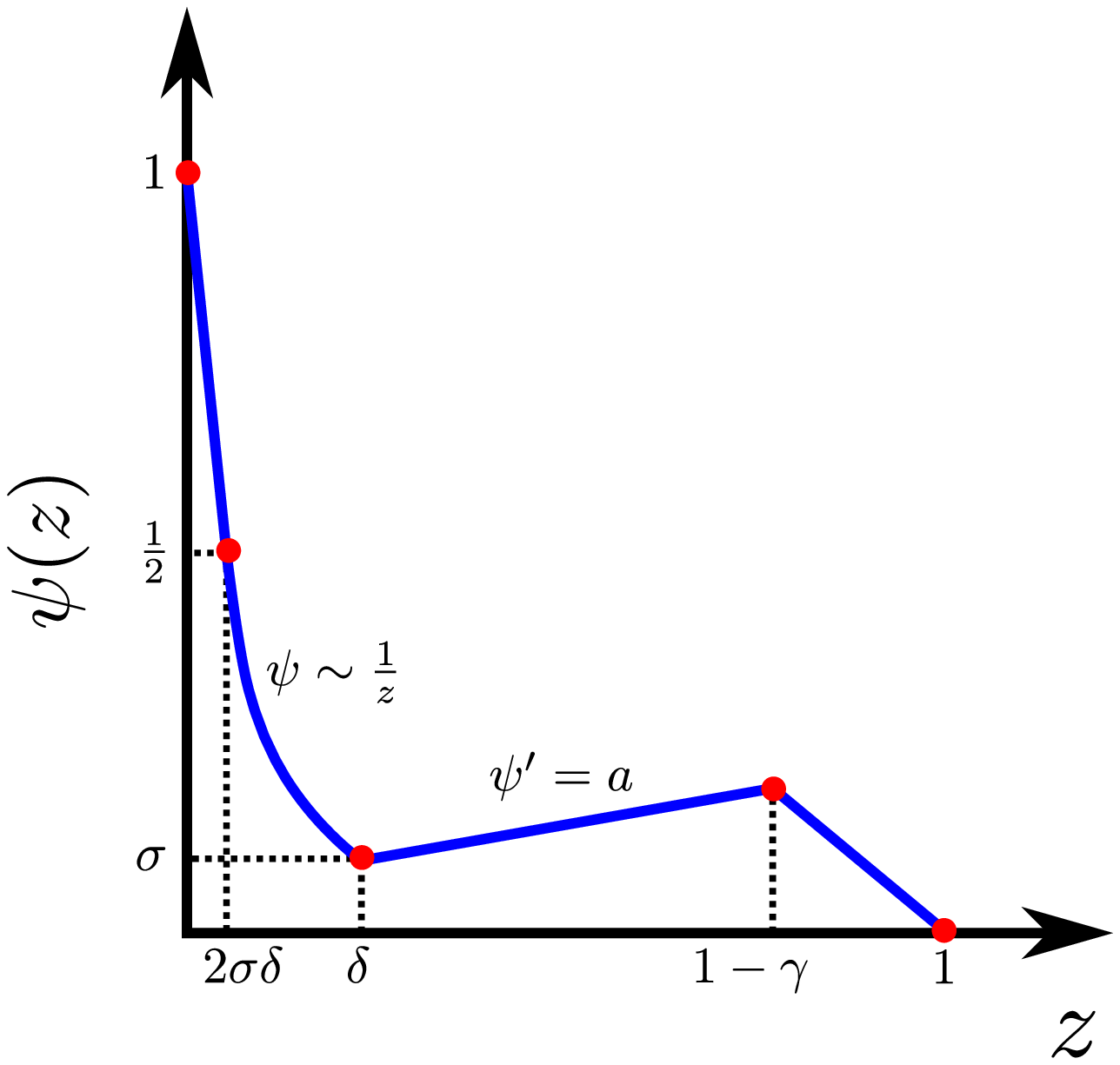}
\caption{}
\end{subfigure} &
\begin{subfigure}{0.5\textwidth}
\centering
\includegraphics[scale = 0.4]{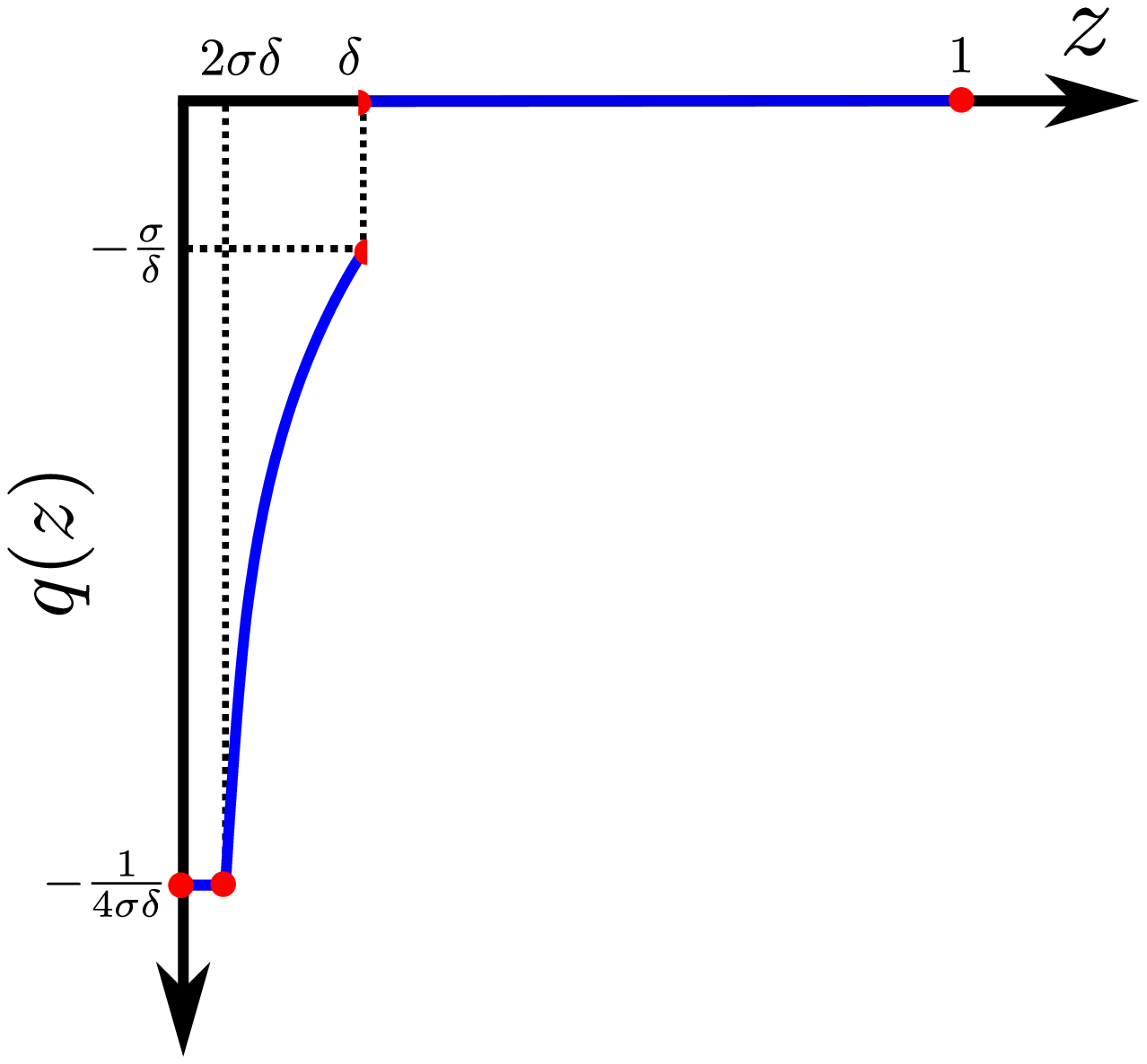}
\caption{}
\end{subfigure}
\end{tabular}
\caption{Sketch of the functions $\psi(z)$ and $q(z)$ from (\ref{Bound in IH1: Def. of psi}), used to obtain a bound on the heat flux $\langle w T \rangle$ in the IH1 configuration.}
\label{Bound in IH1: Profile of psi and q IH1}
\end{figure}

To prove the bound in~\eqref{e:ih1-bound-intro}, we start by setting%making the following choices of the functions $\psi(z)$ and $q(z)$
\begin{eqnarray}
\psi(z) = 
\begin{cases}
1 - \frac{z}{4 \sigma \delta} \quad 0 \leq z \leq 2 \sigma \delta, \\[5pt]
\frac{\sigma \delta}{z} \qquad 2 \sigma \delta \leq z \leq \delta, \\[5pt]
\sigma + a (z - \delta) \quad \delta \leq z \leq 1 - \gamma, \\[5pt]
(1-z) \frac{\sigma + a(1-\gamma-\delta)}{\gamma} \quad 1 - \gamma \leq z \leq 1,
\end{cases} q(z) = 
\begin{cases}
- \frac{1}{4 \sigma \delta} \qquad 0 \leq z \leq 2 \sigma \delta, \\[5pt]
- \frac{\sigma \delta}{z^2} \qquad 2 \sigma \delta \leq z \leq \delta, \\[5pt]
0 \qquad \delta \leq z \leq 1.
\end{cases}
\label{Bound in IH1: Def. of psi}
\end{eqnarray}
These functions are sketched in figure \ref{Bound in IH1: Profile of psi and q IH1}. In the definition of $\psi$, the parameter $\delta$ denotes the thickness of the boundary layer near the bottom plate. The parameter $\sigma$ is the value of $\psi$ taken at the edge of lower boundary layer ($z = \delta$). The lower boundary layer itself is divided into two parts, an inner sublayer where $\psi$ is linear and an outer sublayer where $\psi \sim z^{-1}$. These sublayers meet at an intermediate point ($z=2\sigma\delta$) where both the value and slope of $\psi$ are equal. The inverse-$z$ scaling of $\psi$ in the outer part of the lower boundary layer is one of the key ingredients in proving~\eqref{e:ih1-bound-intro}. The linear inner sublayer, instead, is used to satisfy the boundary condition $\psi(0) = 1$ from (\ref{Formulation: bound on wT constraints}b). In the bulk of the layer ($\delta \leq z \leq 1-\gamma$) we have $\psi^\prime = a$, so the indefinite sign term in (\ref{Formulation: expression Sk}) is zero. Thus, we only need to control the indefinite sign term in the boundary layers. The parameter $\gamma$ is the thickness of the boundary layer near the upper boundary in which the profile of $\psi$ is linear. 

The sole purpose behind the choice of the function $q(z)$ is to ensure $\psi^\prime - q = 0$ in the lower boundary layer, thereby making the positive contribution from the second term in the bound (\ref{Formulation: bound on wT}) small in this layer. All parameters are taken to satisfy 
\begin{eqnarray}
a, b, \sigma, \delta, \gamma \leq 1
\label{Bound in IH1: constraint on a b s d g}
\end{eqnarray}
and this assumption will be implicit in the proof below.

The goal now is to adjust the free parameters $a, b, \sigma, \delta$ and  $\gamma$ such that the spectral constraint (\ref{Formulation: bound on wT constraints}d) is satisfied and, at the same time, the bound (\ref{Formulation: bound on wT}) is as small as possible. We begin by estimating from above the second term in the bound (\ref{Formulation: bound on wT}):
\begin{eqnarray}
\frac{1}{4 b} \int_0^1 (b z - \psi^\prime + q)^2 \textrm{d}z && \leq \frac{1}{2 b} \int_{0}^{1} b^2 z^2 \; \textrm{d}z + \frac{1}{2 b}\left\|\psi^\prime(z) - q(z) \right\|_2^2 \nonumber \\
&& = \frac{b}{6} + \frac{1}{2 b}\int_{\delta}^{1} |\psi^\prime(z) - q(z)|^2 \; \textrm{d}z \nonumber \\
&& \leq \frac{b}{6} + \frac{1}{b}\int_{\delta}^{1} |\psi^\prime(z)|^2 \; \textrm{d}z + \frac{1}{b}\int_{\delta}^{1} |q(z)|^2 \; \textrm{d}z \nonumber \\
&& \leq \frac{b}{6} + \frac{(\sigma + a)^2}{b \gamma} + \frac{a^2}{b} \nonumber
\\
&&\leq \frac{b}{6} + \frac{2 (\sigma + a)^2}{b \gamma}.
\label{Bound in IH1: bound on the second term}
\end{eqnarray}
Next, we estimate from below the last term in the bound (\ref{Formulation: bound on wT}):
\begin{eqnarray}
\int_{0}^{1} \psi \; {\rm d}z && = \frac{3 \sigma \delta}{2} - \sigma \delta \log(2 \sigma)  + \frac{(2 \sigma + a (1 - \gamma- \delta))}{2} + \frac{(\sigma + a (1 - \gamma - \delta)) \gamma}{2} \nonumber \\
&& \geq - \sigma \delta \log(\sigma).
\label{Bound in IH1: bound on the third term}
\end{eqnarray}
Combining (\ref{Bound in IH1: bound on the second term}) and (\ref{Bound in IH1: bound on the third term}) with (\ref{Formulation: bound on wT}), we obtain
\begin{eqnarray}
\langle w T \rangle \leq \frac{1}{2} + \frac{b}{6} + \frac{2 (\sigma + a)^2}{b \gamma} + \sigma \delta \log(\sigma).
\label{Bound in IH1: bound on wT intermediate 1}
\end{eqnarray}
Assuming that
\addtocounter{equation}{1}
\begin{align}
\tag{\theequation\textit{a,b}}
\frac{b}{6} \leq - \frac{1}{4} \sigma \delta \log(\sigma), \qquad
\frac{2 (\sigma + a)^2}{b \gamma} \leq - \frac{1}{4} \sigma \delta \log(\sigma),
\label{Bound in IH1: bound on wT intermediate 2 inequalities}
\end{align}
which will be the case for the choices of $a$, $b$, $\sigma$, $\delta$, $\gamma$ made below, the right-hand side of~\eqref{Bound in IH1: bound on wT intermediate 1} can be further estimated from above to obtain
\begin{eqnarray}
\langle w T \rangle \leq \frac{1}{2} + \frac{1}{2}\sigma \delta \log(\sigma).
\label{Bound in IH1: bound on wT intermediate 2}
\end{eqnarray}

We now shift our focus to the constraint~(\ref{Formulation: bound on wT constraints}d). Dropping the positive terms proportional to $\vert \hat{w}_{\bs{k}} \vert^2$, $\vert \hat{w}_{\bs{k}}'' \vert^2$ and $\vert \hat{T}_{\bs{k}} \vert^2$, it is enough to verify that
%We use a more restrictive, but simplified, form of the spectral constraint which requires that 
%For instance, with the use of an interpolation inequality 
%\begin{equation}
%    \int^{1}_{0}\frac{|\hat{w}''_{\boldsymbol{k}}|^2}{k^2} + k^2 |\hat{w}_{\boldsymbol{k}}|^2 \,\textrm{d}z \geq \int^{1}_{0} 2|\hat{w}'_{\bs{k}}|^2 \textrm{d}z,
%\end{equation}
%and dropping the term $b k^2 |\hat{T}_{\bs{k}}|$, instead of \eqref{Formulation: expression Sk}, we require
\begin{eqnarray}
\widetilde{\mathcal{S}}(\hat{w}, \hat{T}) \coloneqq \int_{0}^{1} \left[\frac{2 a}{R}|\hat{w}^{\prime}|^2 + b |\hat{T}^{\prime}|^2 - (a - \psi^\prime) \hat{w}\hat{T}\right] {\rm d}z \geq 0.
\label{Bound in IH1: simplified spectral constraint}
\end{eqnarray}   
Here, $\hat{w}$ and $\hat{T}$ satisfy the boundary conditions
\begin{subequations}
\label{Bound in IH1: simplified spectral constraint bc}
\begin{eqnarray}
&& \hat{w}(0) = \hat{w}^\prime(0) = \hat{T}(0) = 0, \\
&& \hat{w}(1) = \hat{w}^\prime(1) = \hat{T}(1) = 0,
\end{eqnarray}
\end{subequations}
where $\hat{w}^\prime(0) = \hat{w}^\prime(1) =0$ is a result of the no-slip boundary condition and the incompressibility of the flow field.
For brevity, we have dropped $\bs{k}$ from the subscript. The positive terms we have dropped could be retained, at the expense of a more complicated algebra, in order to improve various prefactors in the eventual bounds. Since this is not our primary goal and the functional form of the bound one obtains does not change, we work with the stronger constraint~\eqref{Bound in IH1: simplified spectral constraint} to ease the presentation.
%The use of the full spectral constraint (\ref{Formulation: bound on wT constraints}{\color{blue}d}) does not lend any advantage; it is possible to choose admissible profiles $\hat{w}$ and $\hat{T}$ in the full spectral constraint (\ref{Formulation: bound on wT constraints}{\color{blue}d}) such that we require the conditions extracted below (\ref{Bound in IH1: condition from the top boundary} and \ref{Bound in IH1: condition from the bottom boundary}) from this simplified spectral constraint (\ref{Bound in IH1: simplified spectral constraint}) to hold but up to different prefactors. Since the improvement of prefactors is not of primary interest, this justifies our choice of working with the simplified spectral constraint.

Substituting the expression of $\psi$ from (\ref{Bound in IH1: Def. of psi}) into (\ref{Bound in IH1: simplified spectral constraint}) gives
\begin{eqnarray}
\widetilde{S}(\hat{w}, \hat{T}) =  && \int_{0}^{2 \sigma \delta} \left[\frac{2 a}{R}|\hat{w}^\prime|^2 + b|\hat{T}^\prime|^2 - \left(a + \frac{1}{4 \sigma \delta}\right) \hat{w} \hat{T}\right] \; {\rm d} z \nonumber \\ 
&& + \int_{2 \sigma \delta}^{\delta} \left[ \frac{2 a}{R}|\hat{w}^\prime|^2 + b|\hat{T}^\prime|^2 - \left(a + \frac{\sigma \delta}{z^2}\right) \hat{w} \hat{T}\right] \; {\rm d} z \nonumber \\
&& + \int_{1-\gamma}^{1} \left[ \frac{2 a}{R}|\hat{w}^\prime|^2 + b|\hat{T}^\prime|^2 - \left(\frac{\sigma + a (1 - \delta)}{\gamma}\right) \hat{w} \hat{T}\right] \; {\rm d} z.
\label{Bound in IH1: simplified spectral constraint with psi q}
\end{eqnarray}
Since $\widetilde{S}(\hat{w}, \hat{T}) \geq \widetilde{S}(|\hat{w}|, |\hat{T}|)$ with equality when $w$ and $T$ are nonnegative, we shall assume without loss of generality that $\hat{w},\hat{T} \geq 0$. We further observe that, if 
\begin{eqnarray}
8 a \delta \leq \sigma, 
\label{Bound in IH1: extra constraint on d}
\end{eqnarray}
then 
\begin{eqnarray}
&& \frac{9}{2}\frac{\sigma \delta}{(z + \sigma \delta)^2} \geq a + \frac{1}{4 \sigma \delta} \quad \text{when} \quad 0 \leq z \leq 2 \sigma \delta, \nonumber \\
&& \frac{9}{2}\frac{\sigma \delta}{(z + \sigma \delta)^2} \geq a + \frac{\sigma \delta}{z^2} \quad \text{when} \quad 2 \sigma \delta \leq z \leq \delta.
\end{eqnarray}
Assuming that $8 a \delta \leq \sigma$, therefore, we can combine the first two terms in (\ref{Bound in IH1: simplified spectral constraint with psi q}) to conclude
\begin{eqnarray}
\widetilde{S}(\hat{w}, \hat{T}) \geq \widetilde{S}_B(\hat{w}, \hat{T}) + \widetilde{S}_T(\hat{w}, \hat{T})
\label{Bound in IH1: nonnegativity of S}
\end{eqnarray}
where
\begin{subequations}
\label{Bound in IH1: Sl and St}
\begin{eqnarray}
\widetilde{S}_B(\hat{w}, \hat{T}) = \int_{0}^{\delta} \left[\frac{2 a}{R}|\hat{w}^\prime|^2 + b|\hat{T}^\prime|^2 -  \frac{9}{2}\frac{\sigma \delta}{(z + \sigma \delta)^2} \hat{w} \hat{T}\right] \; \textrm{d} z, \\
\widetilde{S}_T(\hat{w}, \hat{T}) = \int_{1-\gamma}^{1} \left[ \frac{2 a}{R}|\hat{w}^\prime|^2 + b|\hat{T}^\prime|^2 - \frac{(\sigma + a)}{\gamma} \hat{w} \hat{T}\right] \; \textrm{d} z.
\end{eqnarray}
\end{subequations}
Next, we derive conditions that ensure $\widetilde{S}_B(\hat{w}, \hat{T})$ and $\widetilde{S}_T(\hat{w}, \hat{T})$ are individually nonnegative, thereby implying the nonnegativity of $\widetilde{S}(\hat{w}, \hat{T})$.

First, we deal with $\widetilde{S}_T(\hat{w}, \hat{T})$.
Using the boundary conditions (\ref{Bound in IH1: simplified spectral constraint bc}{\color{blue}b}) along with the fundamental theorem of calculus and the Cauchy--Schwarz inequality leads to
\addtocounter{equation}{1}
\begin{equation}
\tag{\theequation\textit{a,b}}
|\hat{w}|^2 \leq (1-z)\, \int_{1-\gamma}^{1}|\hat{w}'|^2 \textrm{d} z, \qquad |\hat{T}|^2 \leq (1-z)\, \int_{1-\gamma}^{1}|\hat{T}'|^2 \textrm{d} z.
\label{Bound in IH1: Cauchy Schwarz for St}
\end{equation}
%\begin{equation}
%\tag{\theequation\textit{a,b}}
%|\hat{w}| \leq \sqrt{1-z}\,\lVert \hat{w}' \rVert_2, \qquad %|\hat{T}| \leq \sqrt{1-z}\, \lVert \hat{T}' \rVert_2.
%\label{Bound in IH1: Cauchy Schwarz for St}
%\end{equation}
Using (\ref{Bound in IH1: Cauchy Schwarz for St}) in the expression (\ref{Bound in IH1: Sl and St}{\color{blue}b}) of $\widetilde{S}_T$, along with the AM--GM inequality, implies that $\widetilde{S}_T \geq 0$ if 
\begin{eqnarray}
\gamma (\sigma + a) \leq 4 \sqrt{\frac{2a b}{R}}.
\label{Bound in IH1: condition from the top boundary}
\end{eqnarray}

A condition for the nonnegativity of $\widetilde{S}_B(\hat{w}, \hat{T})$, instead, can be derived using the Hardy inequality given in Lemma~\ref{th:hardy}. First, using the AM-GM inequality, we write
\begin{eqnarray}
\widetilde{S}_B(\hat{w}, \hat{T}) \geq \int_{0}^{\delta} \left[\frac{2 a}{R}|\hat{w}^\prime|^2 + b|\hat{T}^\prime|^2 -  \frac{9}{4}\frac{\sigma \delta \beta}{(z + \sigma \delta)^2} | \hat{w}|^2 -  \frac{9}{4}\frac{\sigma \delta}{(z + \sigma \delta)^2 \beta} | \hat{T}|^2 \right] \; \textrm{d} z
\label{Bound in IH1: S_B Youngs}
\end{eqnarray}
for some constant $\beta > 0$ to be specified later. Then, we can apply Lemma~\ref{th:hardy} to estimate
\begin{equation}
\tag{\theequation\textit{a,b}}
\int_{0}^{\delta} \frac{|\hat{w}|^2}{(z + \sigma \delta)^2} {\rm d}z \leq 4 \int_{0}^{\delta} |\hat{w}'|^2 {\rm d} z, \qquad \int_{0}^{\delta} \frac{|\hat{T}|^2}{(z + \sigma \delta)^2} {\rm d}z \leq 4 \int_{0}^{\delta} |\hat{T}'|^2 {\rm d} z.
\label{Bound in IH1: S_B Hardy}
\end{equation}
Using (\ref{Bound in IH1: S_B Hardy}), (\ref{Bound in IH1: S_B Youngs}), and choosing 
\begin{eqnarray}
\beta = \sqrt{\frac{2 a}{b R}},
\end{eqnarray}
we conclude that $\widetilde{S}_B(\hat{w}, \hat{T})$ is nonnegative if
\begin{eqnarray}
\sigma \delta \leq \frac{1}{9} \sqrt{\frac{2ab}{R}}.
\label{Bound in IH1: condition from the bottom boundary}
\end{eqnarray}

Given (\ref{Bound in IH1: condition from the top boundary}) and (\ref{Bound in IH1: condition from the bottom boundary}), and the functional forms of \eqref{Bound in IH1: bound on wT intermediate 2 inequalities} with respect to the variables, one can show that the bound (\ref{Bound in IH1: bound on wT intermediate 2}) is optimized when $a$ is proportional to $\sigma$ and $\delta$ is proportional to $\gamma$. For simplicity, therefore, we take $a = \sigma$ and $\delta = \gamma$; we expect that different choices affect only the value of various prefactors appearing in the final bound, but not its functional form or the powers of $R$. %if the inequalities above were optimised, at least semi-numerically, the coefficients in front of the $R$ scaling would differ. For brevity we proceed with the aim of finding only the scaling. 
With these additional simplifications, the constraints (\ref{Bound in IH1: condition from the top boundary}), (\ref{Bound in IH1: condition from the bottom boundary}) and \eqref{Bound in IH1: bound on wT intermediate 2 inequalities} are satisfied if we take
%the required constraints is
\begin{subequations}
\label{Bound in IH1: choice of parameters for the bound}
\begin{align}
&a = \sigma = \exp \left(-2^{\frac{8}{5}} 3^{\frac{8}{5}} R^{\frac{3}{5}}\right), \\ 
&b = 2^{\frac{7}{5}} 3^{\frac{6}{5}} R^{\frac{1}{5}} \exp \left(-2^{\frac{8}{5}} 3^{\frac{8}{5}} R^{\frac{3}{5}}\right), \\
&\delta = \gamma = 2^{\frac{6}{5}}3^{-\frac{7}{5}} R^{-\frac{2}{5}}.
\end{align}
\end{subequations}
These choices satisfy the inequalities (\ref{Bound in IH1: constraint on a b s d g}) and (\ref{Bound in IH1: extra constraint on d}) assumed in our derivation provided that $R \geq 2^{\frac{21}{2}} 3^{-\frac{7}{2}} \approx 30.97$.
We therefore conclude from~\eqref{Bound in IH1: bound on wT intermediate 2} that
\begin{equation}
\langle w T \rangle \leq \frac{1}{2} - 2^{\frac{7}{5}} 3^{\frac{1}{5}} R^{\frac{1}{5}} \exp \left(-2^{\frac{8}{5}} 3^{\frac{8}{5}} R^{\frac{3}{5}}\right) \qquad \forall R \geq 2^{\frac{21}{2}} 3^{-\frac{7}{2}}.
\label{Bound in IH1: bound on wT final final}
\end{equation}

%One interesting point worth noticing is the thicknesses of the boundary layers given in (\ref{Bound in IH1: choice of parameters for the bound}{\color{blue}c}). In the limit of $R \to \infty$, the boundary layer scaling $R^{-\frac{2}{5}}$ is thicker as compared to the prediction from the scaling theory of \citet{spiegel1963generalization} but it is thinner as compared to the prediction from the theory of \citet{malkus1954heat} and \citet{ priestley1954vertical}.

We end this section with two remarks. First, the scaling of the upper boundary layer thickness given by (\ref{Bound in IH1: choice of parameters for the bound}{\color{blue}c}) is stronger (i.e. the boundary layer is thinner) than the scalings $\gamma\sim R^{-1/4}$ and $\gamma\sim R^{-1/3}$ implied by classical \citep{malkus1954heat, priestley1954vertical} and ultimate \citep{spiegel1963generalization} scaling arguments for Rayleigh-B\'{e}rnard convection, respectively \citep[for further details see \S 3 in][]{arslan2021bounds}.
Second, if instead of using the Hardy inequality in (\ref{Bound in IH1: Sl and St}) we had used the Cauchy--Schwarz and AM--GM inequalities, as we did in the upper boundary layer,  then we would have obtained the condition
\begin{eqnarray}
- \frac{9}{2} \sigma \delta \left(\frac{1}{1 + \sigma} + \log \left(\frac{\sigma}{1 + \sigma}\right)\right) \leq \frac{1}{2} \sqrt{\frac{2 a b}{R}}, %\nonumber \\
%\implies - \sigma \delta \log \sigma \lesssim \sqrt{\frac{a b}{R}}.
\label{Bound in IH1: Cauchy Schwarz for Sl loss of log sigma}
\end{eqnarray}
and therefore $\sigma \delta \log \sigma \lesssim \sqrt{a b /R}$. 
This is worse than condition (\ref{Bound in IH1: condition from the bottom boundary}) by a factor of $\log \sigma^{-1}$ and, as a result, no bound on $\langle w T \rangle$ strictly smaller than 1/2 can be obtained beyond a certain Rayleigh number.
%However, if one tries a few test profiles of $\hat{w}$ and $\hat{T}$ that satisfies the boundary conditions (\ref{Formulation: bound on wT constraints}e-f) at $z = 0$, e.g., $\hat{w} = c_1 z^{\alpha_1}$ and $\hat{T} = c_2 z^{\alpha_2}$, then it becomes apparent that the required condition for the nonnegativity of $\widetilde{S}_B(\hat{w}, \hat{T})$ should be
%\begin{eqnarray}
%\sigma \delta \lesssim \sqrt{\frac{a b}{R}}.
%\end{eqnarray}

\section{Bound on heat flux in IH3}
\label{Bound on heat flux in IH3 configuration lala}

\begin{figure}
\centering
\begin{tabular}{lc}
\begin{subfigure}{0.5\textwidth}
\centering
\includegraphics[scale = 0.4]{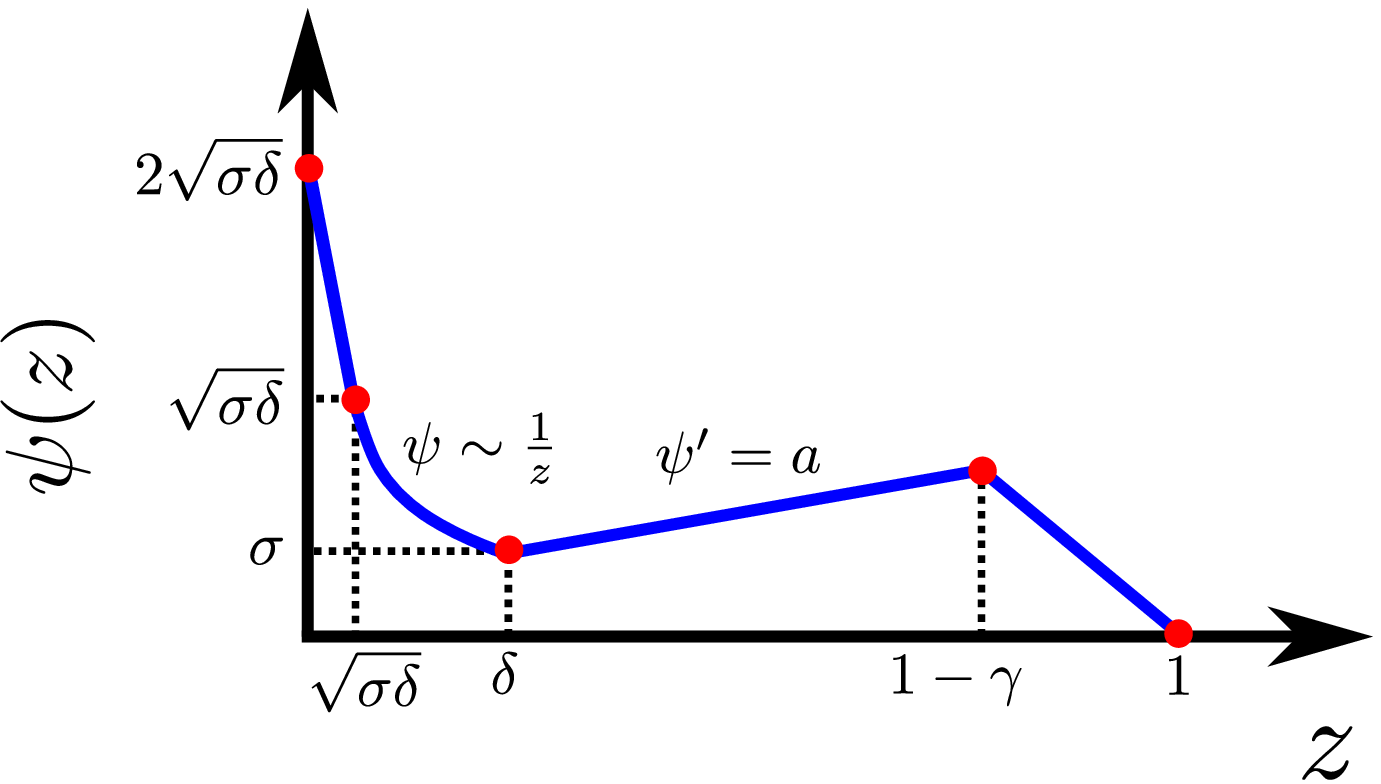}
\caption{}
\end{subfigure} &
\begin{subfigure}{0.5\textwidth}
\centering
\includegraphics[scale = 0.4]{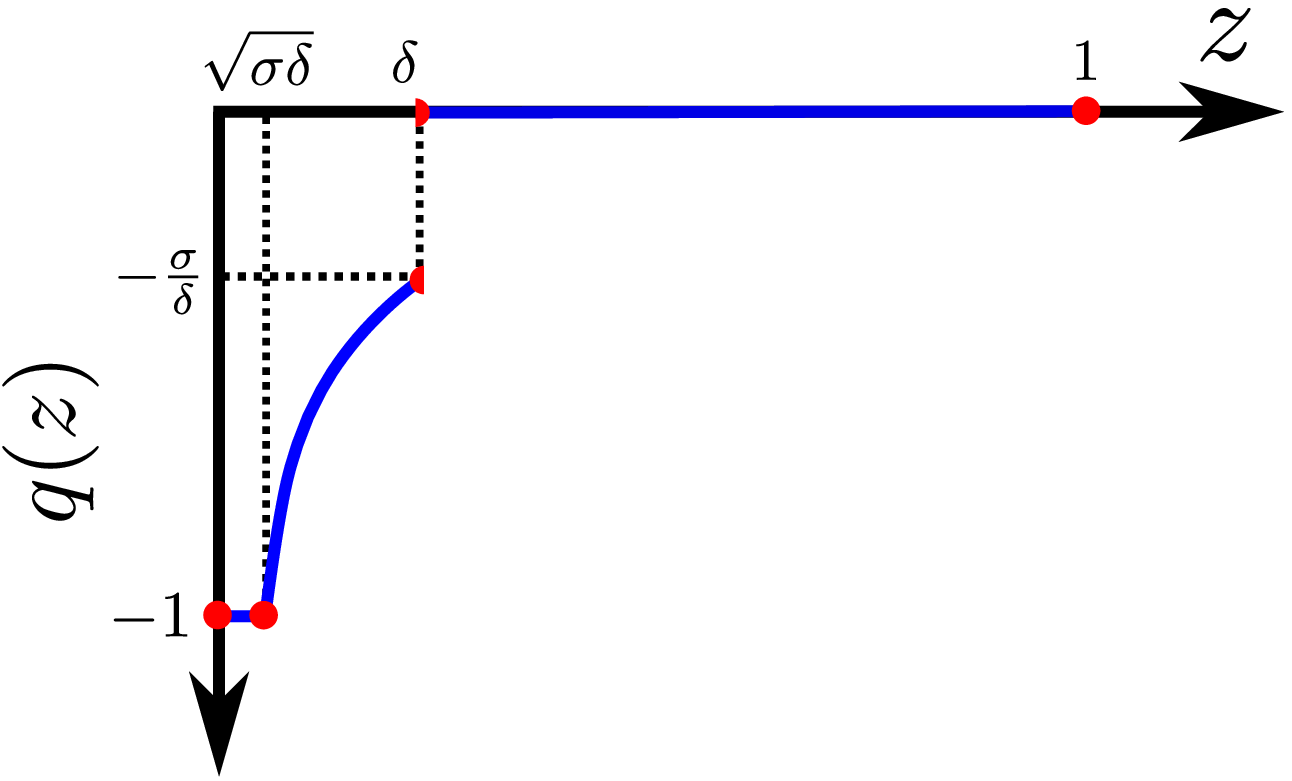}
\caption{}
\end{subfigure}
\end{tabular}
\caption{Sketch of the functions $\psi(z)$ and $q(z)$ from (\ref{Bound in IH3: Def. of psi}), used to obtain bound on the heat flux $\langle w T \rangle$ in the IH3 configuration.}
\label{Bound in IH3: Profile of psi and q IH3}
\end{figure}

We now prove the bound~\eqref{e:ih3-bound-intro} for the IH3 configuration. Similar to the previous section, the key ingredients of the proof are (i) a profile of $\psi$ proportional to $1/z$ near the bottom boundary, and (ii) the use of a nonstandard Rellich inequality. 
We start by choosing the functions $\psi(z)$ and $q(z)$:
\begin{eqnarray}
\psi(z) = 
\begin{cases}
2 \sqrt{\sigma \delta} - z \qquad 0 \leq z \leq \sqrt{\sigma \delta}, \\[5pt]
\frac{\sigma \delta}{z} \qquad \sqrt{\sigma \delta} \leq z \leq \delta, \\[5pt]
\sigma + a (z - \delta) \qquad \delta \leq z \leq 1 - \gamma, \\[5pt]
(1-z) \frac{\sigma + a(1-\gamma-\delta)}{\gamma} \quad 1 - \gamma \leq z \leq 1.
\end{cases} q(z) = 
\begin{cases}
- 1 \qquad 0 \leq z \leq \sqrt{\sigma \delta}, \\[5pt]
- \frac{\sigma \delta}{z^2} \qquad \sqrt{\sigma \delta} \leq z \leq \delta, \\[5pt]
0 \qquad \delta  \leq z \leq 1.
\end{cases}
\label{Bound in IH3: Def. of psi}
\end{eqnarray}
These choices are sketched in figure \ref{Bound in IH3: Profile of psi and q IH3} and the parameters $\sigma, \delta$ and  $\gamma$ have the same purpose as in the last section. The difference between these profiles and those used for the IH1 configuration in \S\ref{Bound on heat flux in IH1 configuration lala} is in the bottom boundary layer ($0 \leq z \leq \delta$). Here, we require $q(0) = -1$ and at the same time want  $q - \psi^\prime = 0$ in the lower boundary. To satisfy these requirements we take the linear boundary sublayer of $\psi$ near the bottom boundary ($0 \leq z \leq \sqrt{\sigma \delta}$) to have slope equal to $-1$. As before, in the outer part of bottom boundary layer ($\sqrt{\sigma \delta} \leq z \leq \delta$),  $\psi$ behaves like $z^{-1}$ and matches smoothly with inner part up to the first derivative. At the edge of the bottom boundary layer ($z = \delta$), the value of $\psi$ is  $\sigma$. In the proof below, we assume
\begin{eqnarray}
a, b, \sigma, \delta, \gamma \leq 1
\label{Bound in IH3: constraint on a b s d g}
\end{eqnarray}

Estimating the second term in the bound (\ref{Formulation: bound on wT}) from above gives
\begin{eqnarray}
\frac{1}{4 b} \int_0^1 (b z - \psi^\prime + q)^2 {\rm d}z && \leq \frac{b}{6} + \frac{2 (\sigma + a)^2}{b \gamma},
\label{Bound in IH3: bound on the second term}
\end{eqnarray}
while the last term can be estimated from below as
\begin{eqnarray}
\int_{0}^{1} \psi \; {\rm d}z && \geq - \frac{1}{2}\sigma \delta \log \left(\frac{\sigma}{\delta}\right).
\label{Bound in IH3: bound on the third term}
\end{eqnarray}
Combining (\ref{Bound in IH3: bound on the second term}) and (\ref{Bound in IH3: bound on the third term}) with (\ref{Formulation: bound on wT}), we obtain
\begin{eqnarray}
\langle w T \rangle \leq \frac{1}{2} + \frac{b}{6} + \frac{2 (\sigma + a)^2}{b \gamma} + \frac{1}{2}\sigma \delta \log \left(\frac{\sigma}{\delta}\right).
\label{Bound in IH3: bound on wT intermediate 1}
\end{eqnarray}
Finally, we assume that 
\begin{eqnarray}
\frac{b}{6} \leq - \frac{1}{8} \sigma \delta \log \left(\frac{\sigma}{\delta}\right), \qquad 
\frac{2 (\sigma + a)^2}{b \gamma} \leq - \frac{1}{8} \sigma \delta \log \left(\frac{\sigma}{\delta}\right)
% && (\sigma + a) \gamma \leq 4 \sqrt{\frac{2 a}{R}}, \\
% && \sigma \delta \leq \frac{1}{3}\sqrt{\frac{a b}{R}}.
\label{Bound in IH3: bound on wT intermediate 2 inequalities}
\end{eqnarray}
(these constraints will be verified later) and estimate the right-hand side of \eqref{Bound in IH3: bound on wT intermediate 1} to arrive at the simpler bound
\begin{eqnarray}
\langle w T \rangle \leq \frac{1}{2} + \frac{1}{4}\sigma \delta \log \left(\frac{\sigma}{\delta}\right).
\label{Bound in IH3: bound on wT intermediate 2}
\end{eqnarray}

For this bound to be valid, we need to adjust the parameters $a$, $b$, $\delta$, $\gamma$ and $\sigma$ such that the spectral condition (\ref{Formulation: bound on wT constraints}d) is satisfied. Dropping the positive terms proportional to $\vert \hat{w}_{\bs{k}} \vert^2$, $\vert \hat{w}_{\bs{k}}' \vert^2$ and $\vert \hat{T}_{\bs{k}}' \vert^2$, we will verify the stronger inequality
\begin{eqnarray}
\widetilde{\mathcal{S}}(\hat{w}, \hat{T}) \coloneqq \int_{0}^{1} \left[\frac{a}{Rk^2}|\hat{w}^{\prime \prime}|^2 + b k^2 |\hat{T}|^2 - (a - \psi^\prime) \hat{w}\hat{T}\right] {\rm d}z \geq 0
\label{Bound in IH3: simplified spectral constraint}
\end{eqnarray}   
for all $z$-dependent functions $\hat{w}$ and $\hat{T}$ satisfying the boundary conditions
\begin{subequations}
\label{Bound in IH3: simplified spectral constraint bc}
\begin{eqnarray}
&& \hat{w}(0) = \hat{w}^\prime(0) = \hat{T}'(0) = 0, \\
&& \hat{w}(1) = \hat{w}^\prime(1) = \hat{T}(1) = 0.
\end{eqnarray}
\end{subequations}
Again, we have dropped the subscript $\bs{k}$  to lighten the notation.

%As $\hat{T} (z)$ need not be zero at $z = 0$ in IH3, therefore, it would be impossible to control the indefinite term without retaining the term $b k^2 |\hat{T}|^2$ in the spectral constraint. However, even then, there is a danger to lose control over this indefinite term if $k \ll 1$, therefore, we also retain the term with $1/k^2$ in the coefficient.

Using arguments similar to those used in \S \ref{Bound on heat flux in IH1 configuration lala} and noticing that if 
\begin{eqnarray}
8 a \delta \leq \sigma
\label{Bound in IH3: extra constraint on d}
\end{eqnarray}
then
\begin{eqnarray}
&&  \frac{9}{2}\frac{\sigma \delta}{(z + \sqrt{\sigma \delta})^2} \geq a + 1 \quad \text{when} \quad 0 \leq z \leq \sqrt{\sigma \delta}, \\
&&  \frac{9}{2}\frac{\sigma \delta}{(z + \sqrt{\sigma \delta})^2} \geq a + \frac{\sigma \delta}{z^2} \quad \text{when} \quad \sqrt{\sigma \delta} \leq z \leq \delta,
\label{Bound in IH3: combine lower boundary spectral constraint}
\end{eqnarray}
we can write
\begin{eqnarray}
\widetilde{S}(\hat{w}, \hat{T}) \geq \widetilde{S}_B(\hat{w}, \hat{T}) + \widetilde{S}_T(\hat{w}, \hat{T}),
\label{Bound in IH3: nonnegativity of S}
\end{eqnarray}
where
\begin{subequations}
\label{Bound in IH3: Sl and St}
\begin{eqnarray}
\widetilde{S}_B(\hat{w}, \hat{T}) = \int_{0}^{\delta} \left[\frac{a}{Rk^2}|\hat{w}^{\prime \prime}|^2 + b k^2 |\hat{T}|^2 -  \frac{9}{2} \frac{\sigma \delta}{(z + \sqrt{\sigma \delta})^2} \hat{w} \hat{T}\right] \; \textrm{d} z, \\
\widetilde{S}_T(\hat{w}, \hat{T}) = \int_{1-\gamma}^{1} \left[\frac{a}{Rk^2}|\hat{w}^{\prime \prime}|^2 + b k^2 |\hat{T}|^2 - \frac{(\sigma + a)}{\gamma} \hat{w} \hat{T}\right] \; \textrm{d} z.
\label{Bound in IH3: def Sl St}
\end{eqnarray}
\end{subequations}
Finding a condition under which $\widetilde{S}_T(\hat{w}, \hat{T}) \geq 0$ is straightforward. Using the fundamental theorem of calculus, the boundary conditions on $\hat{w}$ and Cauchy--Schwarz inequality, we obtain
\begin{eqnarray}
|\hat{w}|^2 \leq \frac{4(1-z)^{3}}{9} \int_{1-\gamma}^1 |\hat{w}''|^2 {\rm d}z. 
\label{Bound in IH3: Cauchy Schwarz for St}
\end{eqnarray}
%\begin{eqnarray}
%|\hat{w}| \leq (1-z)^{3/2} \big\lVert \hat{w}'' \big\rVert_2
%\label{Bound in IH3: Cauchy Schwarz for St}
%\end{eqnarray}
Then, substituting (\ref{Bound in IH3: Cauchy Schwarz for St}) in (\ref{Bound in IH3: Sl and St}b) and using the AM-GM inequality shows that $\widetilde{S}_T(\hat{w}, \hat{T})$ is nonnegative as long as
\begin{eqnarray}
(\sigma + a) \gamma \leq 6 \sqrt{\frac{ a b}{R}}.
\label{Bound in IH3: condition from the top boundary}
\end{eqnarray}

To show that $\widetilde{S}_B(\hat{w}, \hat{T})$ is nonnegative, instead, we rely on the Rellich inequality stated in Lemma~\ref{th:rellich}. First, using the AM-GM inequality we estimate
\begin{eqnarray}
\widetilde{S}_B(\hat{w}, \hat{T}) \geq \int_{0}^{\delta} \left[\frac{a}{Rk^2}|\hat{w}^{\prime \prime}|^2 + b k^2 |\hat{T}|^2 -  \frac{9}{4} \frac{\sigma \delta \beta}{(z + \sqrt{\sigma \delta})^4} |\hat{w}|^2 -  \frac{9}{4} \frac{\sigma \delta}{\beta} |\hat{T}|^2\right] \; \textrm{d} z ,
\label{Bound in IH3: SB AM GM}
\end{eqnarray}
for a the positive constant $\beta$ to be specified below. Next, using Lemma~\ref{th:rellich} we obtain
\begin{eqnarray}
\int_{0}^{\delta} \frac{|\hat{w}|^2}{(z + \sqrt{\sigma \delta})^4} {\rm d}z \leq \frac{16}{9} \int_{0}^{\delta} |\hat{w}''|^2 {\rm d} z.
\label{Bound in IH3: SB Rellich}
\end{eqnarray}
Combining (\ref{Bound in IH3: SB Rellich}) in (\ref{Bound in IH3: SB AM GM}) and setting 
\begin{eqnarray}
\beta = \frac{3}{4 k^2} \sqrt{\frac{a}{b R}}
\end{eqnarray}
we conclude that $\widetilde{S}_B(\hat{w}, \hat{T})$ is nonnegative if
\begin{eqnarray}
\sigma \delta \leq \frac{1}{3} \sqrt{\frac{a b}{R}}.
\label{Bound in IH3: condition from the bottom boundary}
\end{eqnarray}

At this stage, all that remains is to choose values for $a$, $b$, $\delta$, $\gamma$ and $\sigma$ such that \eqref{Bound in IH3: bound on wT intermediate 2 inequalities}, \eqref{Bound in IH3: condition from the top boundary} and \eqref{Bound in IH3: condition from the bottom boundary} hold, at least for sufficiently large Rayleigh numbers, while minimizing the right-hand side of~\eqref{Bound in IH3: bound on wT intermediate 2}. For the same reasons explained at the end of \S\ref{Bound on heat flux in IH1 configuration lala}, we simplify the algebra by choosing
$a = \sigma$ and $\delta = \gamma$. Then, optimizing the bound (\ref{Bound in IH3: bound on wT intermediate 2}) subject to \eqref{Bound in IH3: condition from the top boundary} and \eqref{Bound in IH3: condition from the bottom boundary} leads to
\begin{subequations}
\begin{align}
&a = \sigma = \frac{2^{\frac{4}{5}}}{3^{\frac{3}{5}}} \frac{1}{R^{\frac{2}{5}}} \exp \left(-2^{\frac{14}{5}} 3^{\frac{2}{5}} R^{\frac{3}{5}}\right), \\
&b = \frac{2^{\frac{12}{5}}3^{\frac{1}{5}}}{R^{\frac{1}{5}}} \exp \left(-2^{\frac{14}{5}} 3^{\frac{2}{5}} R^{\frac{3}{5}}\right), \\
&\delta = \gamma = \frac{2^{\frac{4}{5}}}{3^{\frac{3}{5}}} \frac{1}{R^{\frac{2}{5}}}.
\end{align}
\end{subequations}
These choices satisfy the constraints in\eqref{Bound in IH3: bound on wT intermediate 2 inequalities} assumed in our proof for all $R\geq 2^{\frac{19}{2}} 3^{-\frac{3}{2}} \approx 139.35$. Thus, from~\eqref{Bound in IH3: bound on wT intermediate 2} we obtain
\begin{eqnarray}
\langle w T \rangle \leq \frac{1}{2} - \frac{2^{\frac{12}{5}}}{3^{\frac{4}{5}}} \frac{1}{R^{\frac{1}{5}}} \exp \left(-2^{\frac{14}{5}} 3^{\frac{2}{5}} R^{\frac{3}{5}}\right) \quad \forall R\geq 2^{\frac{19}{2}} 3^{-\frac{3}{2}}.
\label{Bound in IH3: bound on wT final final}
\end{eqnarray}
% where the constraints on parameters (\ref{Bound in IH3: constraint on a b s d g}) and (\ref{Bound in IH3: extra constraint on d}) require
% \begin{eqnarray}
% R \geq 2^{\frac{19}{2}} 3^{-\frac{3}{2}} \approx 139.35,
% \end{eqnarray}
% for the bound to hold.
%
It is interesting to note that only the boundary layer thicknesses $\delta$ and $\gamma$ have the same $O(R^{-\frac25})$ scaling as for the IH1 configuration. The parameters $\sigma, a, b$ and the correction to $1/2$ in the bound (\ref{Bound in IH3: bound on wT final final}), instead, are all $O(R^{\frac{2}{5}})$ smaller than their corresponding values for the IH1 case.

\section{Discussion and concluding remarks}
\label{Discussion and concluding remarks updated}
We considered the problem of uniform internally heated convection between two parallel boundaries where either both the boundaries are held at the same constant temperature (IH1 configuration) or the temperature at the top boundary is fixed and the bottom boundary is insulating (IH3 configuration). For both configurations we obtained rigorous $R$-dependent bounds on the heat flux using the background method, which we formulated in terms of a quadratic auxiliary function and augmented with a minimum principle that enables one to consider only nonnegative temperature fields in the optimization problem for the bound.
In each configuration, we were able to prove that $\langle w T\rangle < 1/2$ with exponentially decaying corrections. The two essential ingredients in our proofs were a boundary layer with inverse-$z$ scaling in the background field and the use of Hardy and Rellich inequalities, which allow for a refined analysis of the spectral constraint compared to standard Cauchy--Schwarz inequalities. Without any of these two components, the proof breaks down and it appears impossible to obtain $R$-dependent corrections to the uniform $\langle w T \rangle \leq 1/2$ at arbitrarily large Rayleigh numbers.

The exponential rate at which our analytical bounds (\ref{Bound in IH1: bound on wT final final}) and (\ref{Bound in IH3: bound on wT final final}) approach $1/2$ is not inconsistent with the numerically optimal bounds computed by \citet{arslan2021bounds} for the IH1 configuration. These numerical bounds also approach $1/2$ from below rapidly as $R \rightarrow \infty$ and appear to do so faster than any power law, suggesting that the best possible bounds provable with the background method may indeed have the functional form% indicating the possibility of the form of optimal bounds (provable with the classical background method) being
\begin{subequations}
\begin{eqnarray}
\langle w T \rangle \leq \frac{1}{2} -  c_1 R^{\alpha} \exp \left(-c_2 R^{\beta}\right) \quad \text{in IH1,} \\
\langle w T \rangle \leq \frac{1}{2} -  \frac{c_3}{R^{\alpha}} \exp \left(-c_4 R^{\beta}\right) \quad \text{in IH3.}
\end{eqnarray}
\label{Discussion and concluding remarks updated: expected bounds}
\end{subequations}
for some positive exponents $\alpha,\beta$ and positive constants $c_1, c_2, c_3,c_4$. Unfortunately, the range of Rayleigh numbers spanned by the available numerical results does not permit a confindent estimation of these parameters, so we cannot say whether the exponents $\alpha=1/5$ and $\beta=3/5$ of our analytical bounds are or not optimal.

In the case of IH3, if (\ref{Discussion and concluding remarks updated: expected bounds}b) is the correct scaling of the optimal bound in the framework of quadratic auxiliary functions, then we note that it will not be trivial to prove the conjecture \citep[p. 17]{goluskin2016internally}
\begin{eqnarray}
\langle w T \rangle \leq \frac{1}{2} - \frac{C}{R^{1/3}}.
\end{eqnarray}
%contrary to the expectation of \citet[p. 17]{goluskin2016internally} that this bound is likely provable. 
For the IH3 configuration, moreover, any bound on $\langle w T \rangle$ can be translated into a bound on the Nusselt number---defined as the ratio of the mean total heat flux to the conductive heat flux---via the identity 
\begin{equation}
    Nu = \frac{1}{1-2\langle wT\rangle}.
    \label{e:nusselt_ih3}
\end{equation}
In particular,~\eqref{Bound in IH3: bound on wT final final} implies
\begin{equation}
Nu \leq \frac{3^{\frac45}}{2^{\frac{17}{5}}} R^{\frac15} \exp \left(2^{\frac{14}{5}} 3^{\frac25} R^{\frac35} \right).
\end{equation}
The exponential growth of this bound is in stark contrast with the power-law bounds available for Raleigh-B\'enard convection, most of which can be obtained with much simpler arguments that those used here for IH3.
%can be compared to equivalent results for Raleigh-B\'enard convection, where rigorous bounds on $Nu$ with Rayleigh number is of particular interest. In IH3, the Nusselt number is defined by \eqref{e:nusselt_ih3} and for a bound (\ref{Discussion and concluding remarks updated: expected bounds}{\color{blue}b}) on $\langle w T \rangle$, the corresponding bound on $Nu$ is given by  
% \begin{equation}
% Nu \leq \frac{1}{2 c_3} R^{\frac15}\exp \left(c_4 R^{\frac{3}{5}}\right).
% \end{equation}
% We see that in contrast with the standard Rayleigh--B\'enard convection, where it is possible to obtain a bound with power-law scaling, in case of IH3, obtaining a bound even with exponential scaling proved to be remarkably difficult.  

In the case of IH1, we can compare our bound on $\langle w T \rangle$ with 3D direct numerical simulations by \citep{goluskin2015penetrative}, which suggest
\begin{eqnarray}
\langle w T \rangle \sim \frac{1}{2} - \frac{0.8}{R^{0.055}}.
\end{eqnarray}
Again, this slow power-law correction to the asymptotic value of 1/2 contrasts the exponential behaviour of our bound (\ref{Discussion and concluding remarks updated: expected bounds}a). It remains to be seen if this result is truly overly conservative, as one may expect based on phenomenological arguments~\citep{arslan2021bounds}, or if there exist solutions of the governing equations (\ref{Flow configuration: governing equations}) that saturate it. In that regard, there are two approaches generally used in the Rayleigh--B\'enard convection.  The first one is the study of bulk properties of steady-state solutions bifurcating from the pure conduction state has attracted growing interest in recent years \citep{waleffe2015heat, sondak2015optimal, wen2020steady, wen2020steadynoslip, kooloth2021coherent, motoki2021multi}, and it has been shown that they can transport more heat than turbulence \citep{wen2020steadynoslip}. The second one is the optimal wall-to-wall approach \citep{hassanzadeh2014wall,  tobasco2017optimal, motoki2018optimal,  doering2019optimal, souza2020wall}, which concerns designing incompressible flows with a constraint on the kinetic energy or enstrophy that leads to optimal heat transfer. It would be interesting to conduct similar studies for the two cases of internally heated convection studied in this work.

%In Rayleigh--B\'enard convection, the study of bulk properties of steady-state solutions bifurcating from the pure conduction state has attracted growing interest in recent years \citep{waleffe2015heat, sondak2015optimal, wen2020steady, wen2020steadynoslip, kooloth2021coherent, motoki2021multi}, and it has been shown that they can transport more heat than turbulence \citep{wen2020steadynoslip}. It would be interesting to conduct a similar studies for the two cases of internally heated convection studied in this work.

%Further, it would be interesting to study the effect of nonuniform heating on the bound on vertical transport and find out if there is a heating profile that improves the scaling in the correction over the main term and brings the bound closer to the results from experiments or DNS. This will be a question of future consideration.

\section*{Acknowledgement}
A.K. thanks D. Goluskin for a discussion and providing comments on the paper. A.A. acknowledges funding by the EPSRC Centre for Doctoral Training in Fluid Dynamics across Scales (award number EP/L016230/1). G.F. was supported by an Imperial College Research Fellowship.

\section*{Declaration of interests}
The authors report no conflict of interest.

\appendix
\section{Proof of Hardy and Rellich inequalities}
\label{Proof of Hardy and Rellich inequalities}
%In this appendix, we prove the Hardy and Rellich inequalities stated in Lemmas~\ref{th:hardy} and~\ref{th:rellich}. 
% \subsection{Hardy inequality}
% \begin{customlemma}{1}
% \label{Proof of Hardy and Rellich inequalities: Hardy inequality}
% Set $\epsilon > 0$ and let $f: [0, \infty) \to \mathbb{R}$ be function such that $f, f' \in L^2([0, \infty))$ satisfying the boundary conditions $f(0) = 0$ then 
% \begin{eqnarray}
% \int_{0}^{\alpha} \frac{|f|^2}{(z + \epsilon)^2} {\rm d}z \leq 4 \int_{0}^{\alpha} |f'|^2 {\rm d} z
% \end{eqnarray}
% holds for any $\alpha \geq 0$.
% \end{customlemma}
% \proof
% We consider the following transformation
% \begin{eqnarray}
% g(z) = \frac{f(z)}{\sqrt{z + \epsilon}}
% \label{Hardy inequality: a simple transformation}
% \end{eqnarray}
% Then rewriting the square of the derivatives of $f^\prime$ as
% \begin{eqnarray}
%      |f^\prime|^2 && = (z+\epsilon)|g'|^2 + \left(\frac12 g^2\right)' + \frac14 (z+\epsilon)^{-1} |g|^2 \nonumber \\
%      && = (z+\epsilon)|g'|^2 + \left(\frac12 g^2\right)' + \frac14 (z+\epsilon)^{-2} |f|^2
%       \label{Hardy inequality: change of variables w}
% \end{eqnarray}
% Finally, integrating in $z$ from $0$ to $\alpha$ leads to the desired result.
% \QEDB
\subsection{Proof of the Hardy inequality in Lemma~\ref{th:hardy}}
Set $f(z) = g(z) \sqrt{z + \epsilon}$ for a suitable function $g(z)$ satisfying $g(0)=0$, and estimate
\begin{eqnarray}
 |f^\prime|^2 && = (z+\epsilon)|g'|^2 + \left(\frac12 g^2\right)' + \frac14 (z+\epsilon)^{-1} |g|^2 \nonumber \\
 && = (z+\epsilon)|g'|^2 + \left(\frac12 g^2\right)' + \frac14 (z+\epsilon)^{-2} |f|^2\nonumber \\
 && \geq \left(\frac12 g^2\right)' + \frac14 (z+\epsilon)^{-2} |f|^2.
  \label{Hardy inequality: change of variables w}
\end{eqnarray}
Upon integrating this inequality in $z$ from $0$ to $\alpha$ and using the boundary condition $g(0)=0$, we find
\begin{align}
\int_0^\alpha |f^\prime(z)|^2 \,{\rm d}z
&\geq \frac12 g(\alpha)^2 + \frac14 \int_0^\alpha (z+\epsilon)^{-2} |f(z)|^2 \, {\rm d}z \nonumber \\
&\geq \frac14 \int_0^\alpha (z+\epsilon)^{-2} |f(z)|^2 \, {\rm d}z,
\end{align}
which is the desired inequality.

\subsection{Proof of the Rellich inequality in Lemma~\ref{th:rellich}}
% \begin{customlemma}{2}
% \label{Proof of Hardy and Rellich inequalities: Rellich inequality}
% Set $\epsilon > 0$ and let $f: [0, \infty) \to \mathbb{R}$ be function such that $f, f', f'' \in L^2([0, \infty))$ satisfying the boundary conditions $f(0) = f'(0) = 0$ then 
% \begin{eqnarray}
% \int_{0}^{\alpha} \frac{|f|^2}{(z + \epsilon)^4} {\rm d}z \leq \frac{16}{9} \int_{0}^{\alpha} |f''|^2 {\rm d} z
% \end{eqnarray}
% holds for any $\alpha \geq 0$.
% \end{customlemma}
Write $f'(z)=\sqrt{z + \epsilon} g(z)$ and $f(z) = (z + \epsilon)^{3/2} h(z)$ for suitable functions $g$ and $h$ satisfying $g(0)=0=h(0)$. 
% \begin{eqnarray}
% g(z) = \frac{f^\prime(z)}{\sqrt{z + \epsilon}}, \qquad h(z) = \frac{f(z)}{(z + \epsilon)^{3/2}}.
% \label{Rellich inequality: a useful transformation}
% \end{eqnarray}
Then,
\begin{subequations}
\begin{align}
|f^{\prime \prime}|^2 & = (z + \epsilon) |g^\prime|^2 + \frac{g^2}{4 (z + \epsilon)} + \left(\frac{1}{2} g^2 \right)^{\prime} \nonumber \\
&= (z + \epsilon) |g^\prime|^2 + \frac{|f^\prime|^2}{4 (z + \epsilon)^2} + \left(\frac{1}{2} g^2 \right)^{\prime}\nonumber \\
&\geq \frac{|f^\prime|^2}{4 (z + \epsilon)^2} + \left(\frac{1}{2} g^2 \right)^{\prime}
\label{rellich-proof-1}
\intertext{and}
|f^{\prime}|^2 &= (z + \epsilon)^3 |h^\prime|^2 + \frac{9}{4}(z + \epsilon) h^2 + (z + \epsilon)^2 \left(\frac{3}{2} h^2 \right)^{\prime} \nonumber \\
& = (z + \epsilon)^3 |h^\prime|^2 + \frac{9}{4} \frac{\vert f \vert^2}{(z + \epsilon)^{2}} +  (z + \epsilon)^2 \left(\frac{3}{2} h^2 \right)^{\prime}\nonumber \\
& \geq \frac{9}{4} \frac{\vert f \vert^2}{(z + \epsilon)^{2}} +  (z + \epsilon)^2 \left(\frac{3}{2} h^2 \right)^{\prime}
\label{rellich-proof-2}
\end{align}
\end{subequations}
Combining \eqref{rellich-proof-2} and \eqref{rellich-proof-1} and then integrating in $z$ from $0$ to $\alpha$ yields
\begin{align}
\int_0^\alpha |f^{\prime \prime}|^2 {\rm d}z
&\geq \int_0^\alpha \frac{9\vert f \vert^2}{16 (z + \epsilon)^4} + \left(\frac{3}{8} h^2 \right)^{\prime} + \left(\frac{1}{2} g^2 \right)^{\prime} {\rm d}z \nonumber
\\
&= \int_0^\alpha \frac{9\vert f \vert^2}{16 (z + \epsilon)^4} \, {\rm d}z + \frac{3}{8} h(\alpha)^2 + \frac{1}{2} g(\alpha)^2 \nonumber
\\
&\geq \int_0^\alpha \frac{9\vert f \vert^2}{16 (z + \epsilon)^4} {\rm d}z ,
\end{align}
which completes the proof.
%\textbf{[GF: What about the last term here?]}
\QEDB

\bibliographystyle{jfm}
\bibliography{Reference.bib}

\end{document}